\documentclass[amsmath,showpacs,aps,prl,reprint]{revtex4-1}
\pdfoutput=1			% Ensuring PDFLaTeX use on arXiv
\usepackage{bm}
\usepackage{graphics}
\usepackage{graphicx}
\usepackage{amsthm}
\usepackage{amsmath}
\usepackage{wasysym}
\usepackage{amssymb}
\usepackage{enumerate}
\usepackage{xfrac}
\usepackage[dvips]{epsfig}
\usepackage[caption=false]{subfig}
\usepackage[normalem]{ulem}
\usepackage{color}
\usepackage{txfonts}
\usepackage[mathscr]{euscript}
\usepackage{hyperref}	%Consider using this package to place links to references in PDF
\usepackage{verbatim}
\usepackage{gensymb}
\usepackage{ulem}

\begin{document}
\title{Topological Magnon Bands and Unconventional Superconductivity in Pyrochlore Iridate Thin Films}
\author{Pontus Laurell}
\email{laurell@physics.utexas.edu}
\author{Gregory A. Fiete}
\affiliation{Department of Physics, The University of Texas at Austin, Austin, TX 78712, USA}
\date{\today}

%75.25.-j	spin arrangements in magnetically ordered materials (including neutron and spin-polarized electron studies, synchrotron-source x-ray scattering, etc.) 
%75.30.Ds	spin waves
%71.27.+a, 	Strongly correlated electron systems; heavy fermions

%%%%%%%%%%%%%%%%%%%%%%%%%%%%%%%%%%%%%%%%%%%%%%%
% Abstract
%%%%%%%%%%%%%%%%%%%%%%%%%%%%%%%%%%%%%%%%%%%%%%%
\begin{abstract}
We theoretically study the magnetic properties of pyrochlore iridate bilayer and trilayer thin films grown along the $[111]$ direction using a strong coupling approach. We find the ground state magnetic configurations on a mean field level and carry out a spin-wave analysis about them. In the trilayer case the ground state is found to be the all-in-all-out (AIAO) state, whereas the bilayer has a deformed AIAO state. For all parameters of the spin-orbit coupled Hamiltonian we study, the lowest magnon band in the trilayer case has a nonzero Chern number. In the bilayer case we also find a parameter range with nonzero Chern numbers. We calculate the magnon Hall response for both geometries, finding a striking sign change as function of temperature. Using a slave-boson mean-field theory we study the doping of the trilayer system and discover an unconventional time-reversal symmetry broken $d+id$ superconducting state.  Our study complements prior work in the weak coupling limit and suggests that the $[111]$ grown thin film pyrochlore iridates are a promising candidate for topological properties and unconventional orders.  
\end{abstract}
\maketitle

%%%%%%%%%%%%%%%%%%%%%%%%%%%%%%%%%%%%%%%%%%%%%%%
%\section{Introduction}
%%%%%%%%%%%%%%%%%%%%%%%%%%%%%%%%%%%%%%%%%%%%%%%
\emph{Introduction - }%
One of the main focal points of quantum materials research is topological states of matter, where the essential physics is generally due to spin-orbit coupling (SOC) \cite{RevModPhys.83.1057,RevModPhys.82.3045,Moore:nat10,Ando:jpsj13}. Another is correlated electron systems, in which electron-electron interactions dominate, leading to  interaction-driven insulators, magnetism, and unconventional superconductivity \cite{RevModPhys.78.17}.  Even more possibilities open up when both SOC and electron-electron interactions are present, such as fractionalized topological insulators (TI) \cite{PhysRevLett.103.196803,Maciejko2015,Maciejko:prl14,Ruegg:prl12,Kargarian:prl13,PhysRevB.83.165112,Maciejko:prb13,Maciejko:prl10,Maciejko:prb12,Stern:arcmp16,Pesin2010,Young:prb08}, interaction-induced TI \cite{PhysRevLett.100.156401,Wen:prb10,Weeks:prb10a,Zhang:prb09,Yang:prb11a,Ruegg11_2,Wang:prb15}, and  unconventional magnetic states \cite{Kargarian:prb12,Reuther:prb12,Rau:prl14}. A promising place to look for both types of physics is in $5d$ transition-metal oxides, which tend to have comparable electron-electron interaction and SOC strengths \cite{Witczak-Krempa2013,Rau:arcm16,Schaffer:cm15}.
	
Among these, the pyrochlore iridates $A_2$Ir$_2$O$_7$ (where $A$ is a rare earth element) have seen a lot of interest \cite{Pesin2010,PhysRevB.85.045124,PhysRevB.86.235129,PhysRevB.87.155101,PhysRevB.83.165112,Maciejko:prl14,Kargarian:prl13,PhysRevB.83.205101,Yang:prb10b}, in part due to the geometrically frustrated lattice suggesting exotic magnetic phenomena such as chiral spin liquids \cite{Machida2010}, and in part since some iridates allow an elegant treatment of the iridium orbitals in the form of a single effective $j_{eff}=1/2$ moment \cite{PhysRevLett.101.076402,Kim2009,Kanamori1957,Zhang_Huale:prl13}. (The splitting of the $t_{2g}$ orbitals into separate $j=1/2$ and $j=3/2$ manifolds may be violated in practice, but still represents a good starting point unless one intends to make quantitatively accurate first-principles predictions \cite{Hu2014,Shinaoka2015}.) Magnetically, the bulk systems tend to order in a noncollinear $q=0$ all-in/all-out (AIAO) configuration \cite{PhysRevB.86.014428,PhysRevB.87.100403,Disseler2014,PhysRevLett.117.037201,Liang2016,Shinaoka2015}, although Pr$_2$Ir$_2$O$_7$ shows no signs of order at the lowest experimentally accessible temperatures \cite{PhysRevLett.96.087204,PhysRevB.94.205107}. As for electronic phases, there are many theoretical possibilities, including axion insulators \cite{PhysRevB.83.205101,PhysRevLett.109.066401,PhysRevB.86.235129}, topological Mott insulators \cite{Pesin2010,PhysRevB.83.165112}, topological crystalline insulators \cite{Kargarian:prl13}, other varieties of correlated topological insulators \cite{Maciejko:prl14}, and Weyl semimetals \cite{PhysRevB.83.205101,PhysRevB.84.075129,PhysRevB.85.045124,PhysRevB.86.235129}. Thin films grown along high symmetry directions have been shown to offer further possible phases, that are not readily inferred from the bulk ones \cite{PhysRevB.86.235141,Hu2014,Chen2015,Yang2014,Hwang2016-Science,Bergholtz2014}.
	
Previous thin film studies have approached the system from the weak coupling limit \cite{Hu2014,Fiete:jap15}, but the materials are really in the non-perturbative intermediate coupling regime \cite{Witczak-Krempa2013,Rau:arcm16,Schaffer:cm15}. Here we describe work starting from the strong-coupling limit, offering a perspective complementary to previous studies. Alternatively, one can view our work as a study of spin (local moment) models with SOC on decorated kagome lattices, possibly artificially created in an optical lattice using artificial gauge fields to realize the SOC and spin interaction terms \cite{RevModPhys.83.1523,PhysRevLett.109.085302,PhysRevLett.109.085303,Gong2015}. While experiments on Eu$_2$Ir$_2$O$_7$ provide evidence for an AIAO order in thicker thin films \cite{Fujita2015,Fujita2016,Fujita2016a}, theoretical studies on atomically thin bilayers and trilayers are less conclusive \cite{PhysRevB.86.235141,Hu2014,Chen2015}. We find that, in the strong coupling limit and on the mean field level, the magnetic ground state is AIAO in the trilayer system, and closely related to it in the bilayer. Performing a spin-wave analysis about the magnetic ground state, we find the lowest energy band is isolated and has a nonzero Chern number.  We calculate the associated (thermal) magnon Hall response, which shows strong signatures of the band topology. Finally we introduce dopants into the system, and explore the phase diagram in a slave-boson mean field treatment, discovering that a large region of parameter space hosts an unconventional time-reversal symmetry broken $d+id$ superconducting phase in which the order parameter sign alternates between layers.

%%%%%%%%%%%%%%%%%%%%%%%%%%%%%%%%%%%%%%%%%%%%%%%
%\section{Magnetic groundstate}
%%%%%%%%%%%%%%%%%%%%%%%%%%%%%%%%%%%%%%%%%%%%%%%
\emph{Spin model - }
The spin Hamiltonian for the bulk system is \cite{PhysRevX.1.021002}
\begin{align}
	H	&=	\sum_{\langle ij\rangle} \left[ J \mathbf{S}_i \cdot \mathbf{S}_j + \mathbf{D}_{ij} \cdot \left( \mathbf{S}_i \times \mathbf{S}_j \right) + S_i^a \Gamma_{ij}^{ab} S_j^b \right],	
	\label{Hmag}
\end{align}
where ${\bf S}_i$ is the effective spin-1/2 moment on site $i$, and the terms represent the antiferromagnetic (AF) Heisenberg coupling, Dzyaloshinskii-Moriya (DM) interaction (DMI) and symmetric anisotropic exchange, respectively. The latter two are given in terms of the normalized DM vectors $\hat{v}_{ij}$ \cite{PhysRevB.85.045124}, by $\mathbf{D}_{ij} = DM \hat{v}_{ij}$ and $\Gamma_{ij}^{ab} =	\Gamma \left( \hat{v}_{ij}^a \hat{v}_{ij}^b - \delta^{ab}/3 \right)$, parametrized by the strengths $DM$ and $\Gamma$, respectively. In the strong coupling limit, the three interaction strengths can be determined microscopically using a Slater-Koster (SK) approach \cite{PhysRevB.85.045124,PhysRevB.87.155101}. In the simplest models they are functions of the SK parameter $t_\sigma$ with an overall scaling due to the Hubbard $U$ coupling \cite{PhysRevB.85.045124,PhysRevB.87.155101}. It is also possible to determine the direction of the DM vectors (up to an overall sign) from symmetry alone \cite{PhysRevB.71.094420,PhysRevLett.4.228,PhysRev.120.91}, producing two configurations known as direct and indirect, associated with an AIAO order and a noncoplanar order, respectively. For bulk pyrochlore iridates, the direct configuration is expected and there is indeed also direct experimental evidence for the AIAO order \cite{Disseler2014}.

\begin{figure}
	\subfloat[][]{
				\raisebox{.8cm}{\includegraphics[width=.3\linewidth]{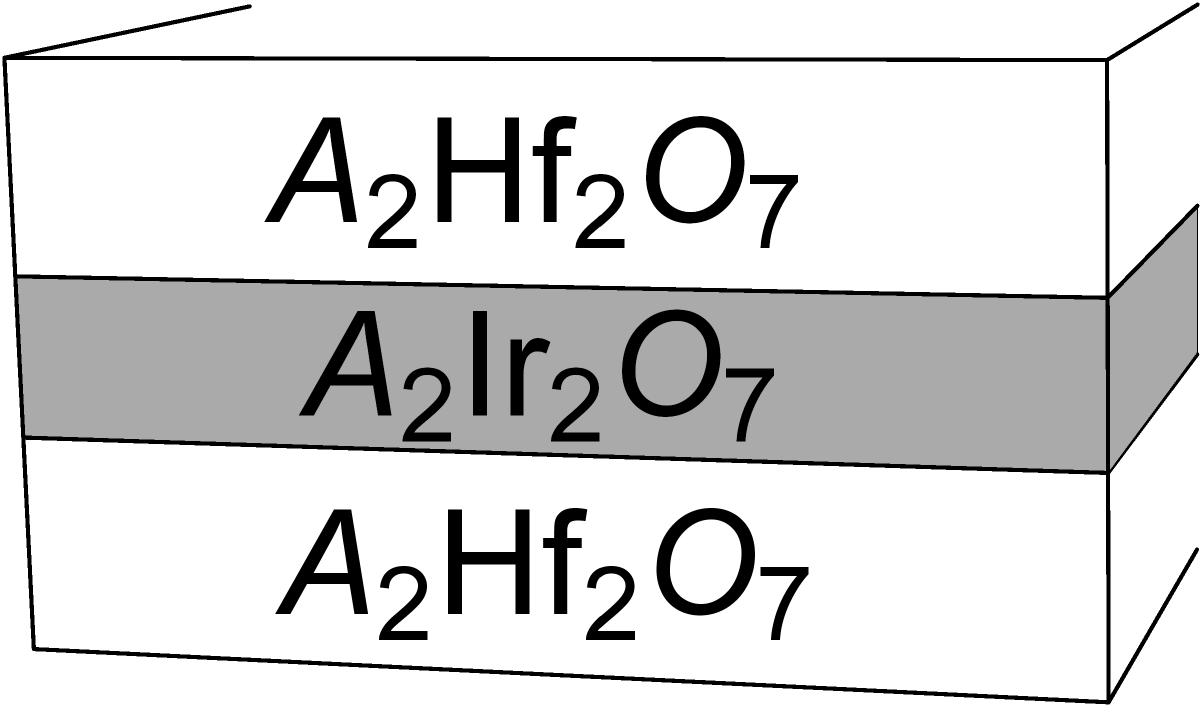}}
				\label{sandwich}
	}
	\subfloat[][TKT lattice]{
				 \includegraphics[height=3cm]{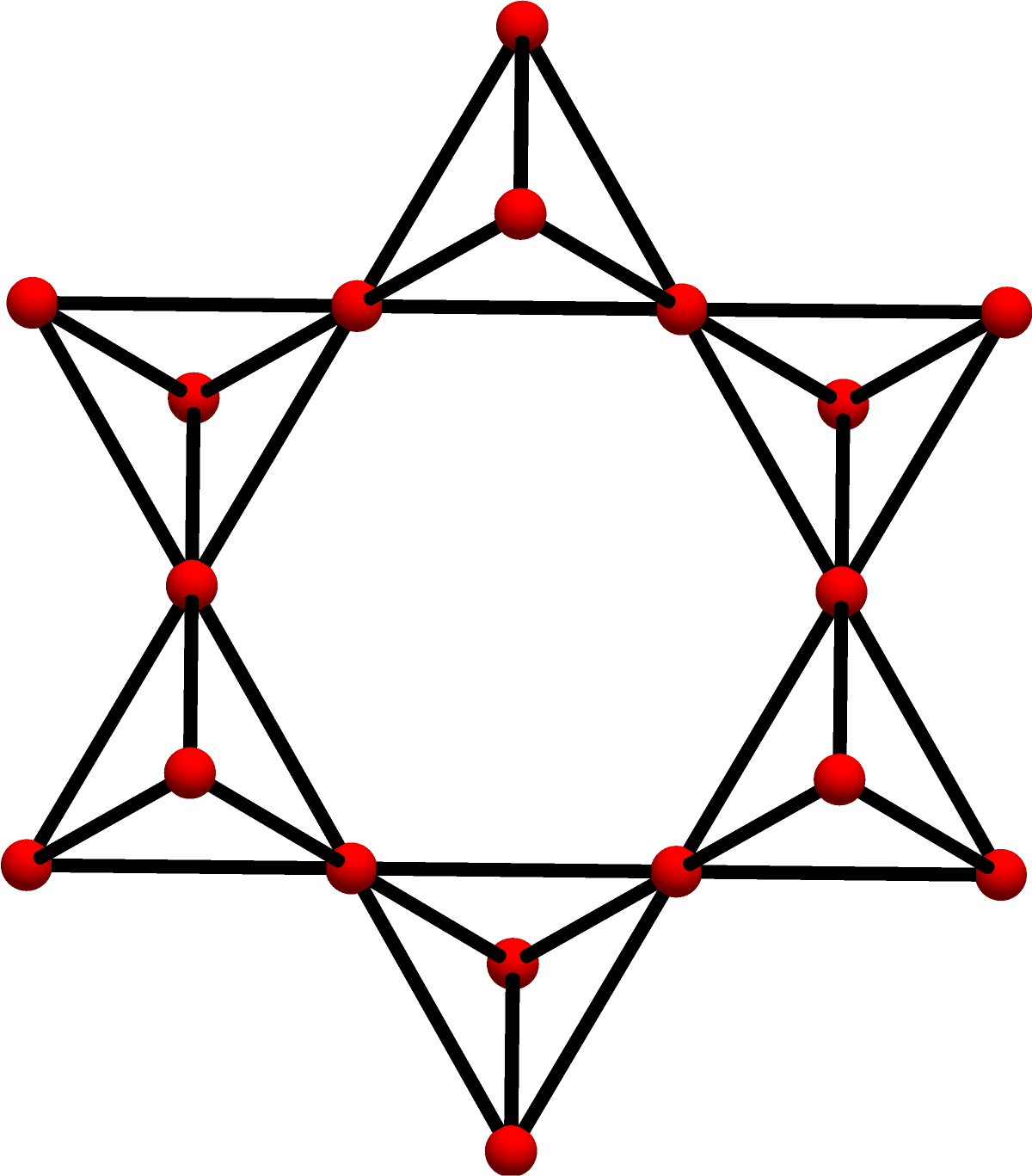}
				\label{TKT_lattice}
	}
	\subfloat[][KT lattice]{
				 \includegraphics[height=3cm]{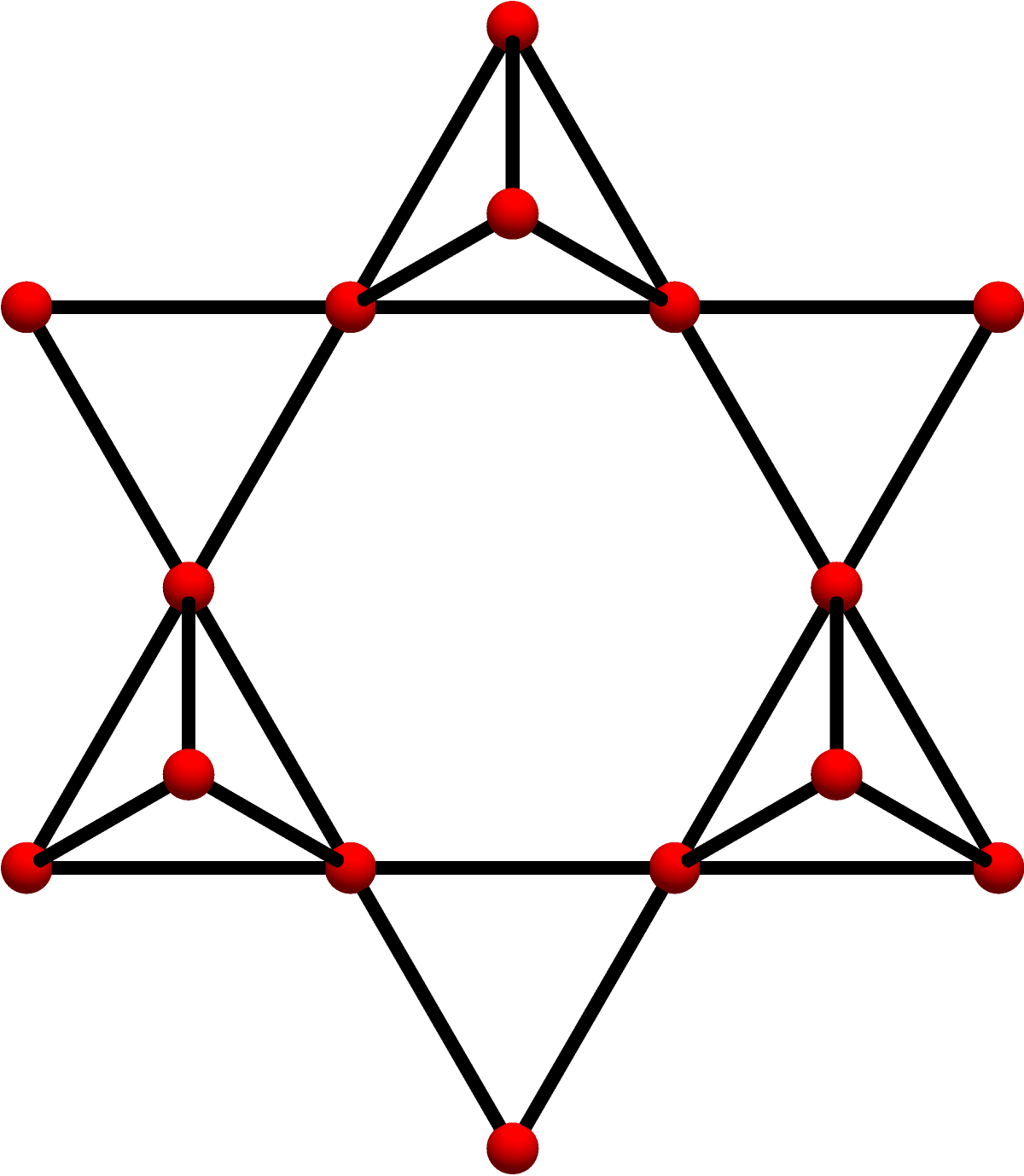}
				\label{KT_lattice}
	}
			\caption{\label{Lattices}(color online) (a) ``Sandwich'' structure. (b) Top view of trilayer grown along [111]. (c) Top view of bilayer structure.}
\end{figure}

In this work we focus on thin films of pyrochlore iridates 
with nonmagnetic A site ions \cite{Supplemental} 
grown \cite{Yang:cm16} in the $[111]$ direction \cite{PhysRevB.86.235141,Hu2014,Chen2015,Yang2014,Hwang2016-Science,Bergholtz2014}. Along this axis, the pyrochlore lattice consists of alternating triangular and kagome layers. We focus on triangular-kagome-triangular (TKT) trilayers and kagome-triangular (KT) bilayers \cite{PhysRevB.86.235141,Hu2014,Chen2015}, as shown in Fig.~\ref{Lattices}. It is assumed that the spin Hamiltonian for these layers is the same as in the bulk case, but that the lattice develops inequivalent sublattices with different coordination numbers.  For example, in the TKT case sites in the triangular (kagome) layers have 3 (6) nearest neighbors (NNs). The absence of a mirror plane means that we are unable to determine the thin film DM vectors from Moriya's rules. Instead, we take them to be inherited from the bulk lattice \cite{Supplemental}. To this end, and to minimize distortion effects, we consider a ``sandwich structure'' with lattice matched support layers, as shown in Fig.~\ref{Lattices}(a).

\begin{figure}
	\subfloat[][TKT mean field state]{
				 \includegraphics[width=.5\linewidth]{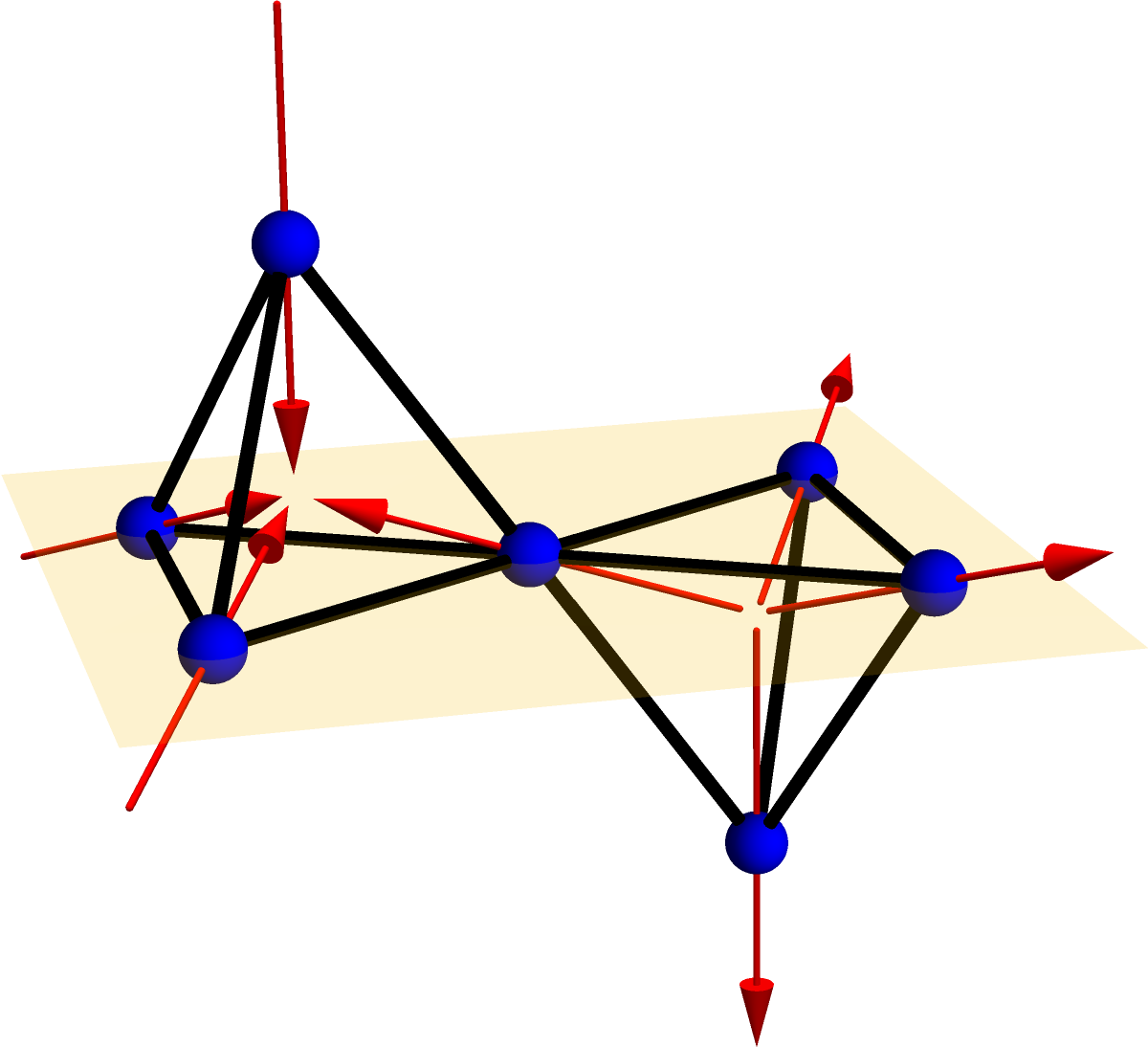}
				\label{MFT_TKT}
	}
	\subfloat[][KT mean field state]{
				 \raisebox{1.25cm}{\includegraphics[width=.5\linewidth]{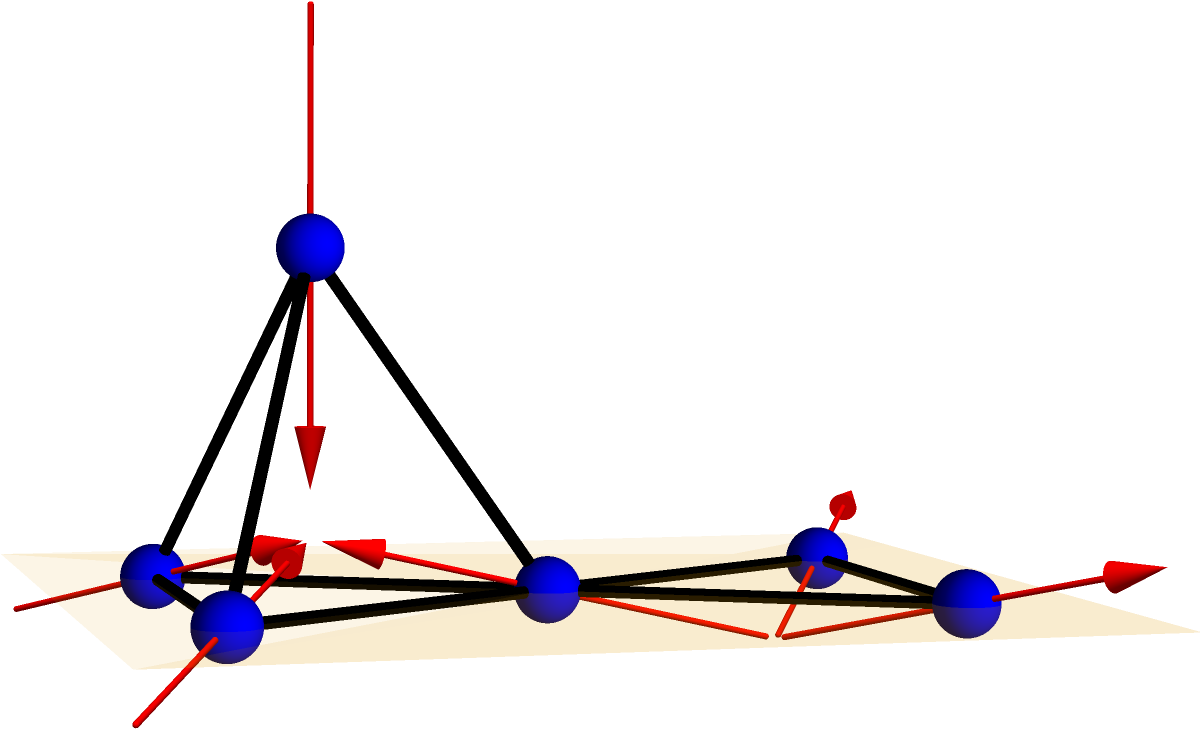}}
				\label{MFT_KT}
				\newline
			}
			\caption{(color online) (a) The AIAO structure generically found for the TKT trilayer, and (b) the deformed AIAO state typical for the KT bilayer. The shaded planes represent the kagome planes.}
\end{figure}

Following bulk results \cite{PhysRevB.71.094420,PhysRevB.83.205101} as well as thin film predictions \cite{PhysRevB.86.235141,Hu2014,Chen2015,Yang2014,Hwang2016-Science}, we assume a $\mathbf{q}=0$ (translationally invariant) structure and solve the model variationally on the mean-field level. In the case of the TKT trilayer, we generically find the AIAO state, as shown in Fig.~\ref{MFT_TKT}. In the KT bilayer case, the symmetry is lowered even further, and we find a family of distorted AIAO states. Essentially, these are spin configurations similar to the AIAO structure, but the spins do not meet in the center of the tetrahedron. The $z$ coordinate of this intersection point can be used to classify the magnetic configuration and measure how far from the AIAO state it is. A typical case is shown in Fig.~\ref{MFT_KT}. However, if we set $J=1$ and choose $DM<0$ and $\Gamma$ freely, eschewing direct ties to microscopic models, we find that it is possible to get arbitrarily close to the AIAO state \cite{Supplemental}. 

Reaching the AIAO state in the bilayer does require a fine-tuning of both $DM$ and $\Gamma$, as well as an unusually high value for the DM strength with $|DM|/J \gtrsim 1$, whereas experiments suggest $|DM|/J \sim 0.1 - 0.3$ \cite{PhysRevLett.117.037201}. 
Nevertheless, it is interesting to consider these bilayer mean field states proximate to the AIAO order for $|DM|/J \ge 0.7$ as they yield topological lowest magnon bands, whereas those found for the parameters used in the TKT case do not. 
In addition, we note that optical lattice systems may allow strengths up to $|DM|/J=1$ \cite{Gong2015}, rendering our work potentially relevant to those experimental systems.

\begin{figure*}[t]
	\subfloat[][TKT spectra]{
				\includegraphics[height=3.6cm]{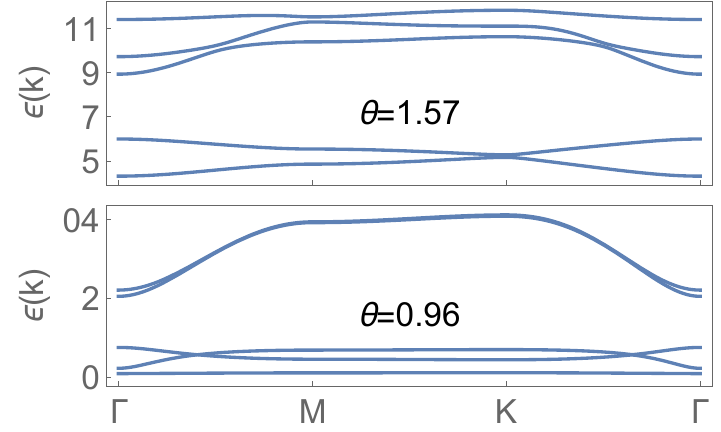}
				\label{MagnonTKT}
	}
	\subfloat[][TKT phase diagram]{
				\raisebox{-0.18cm}{\includegraphics[height=3.6cm]{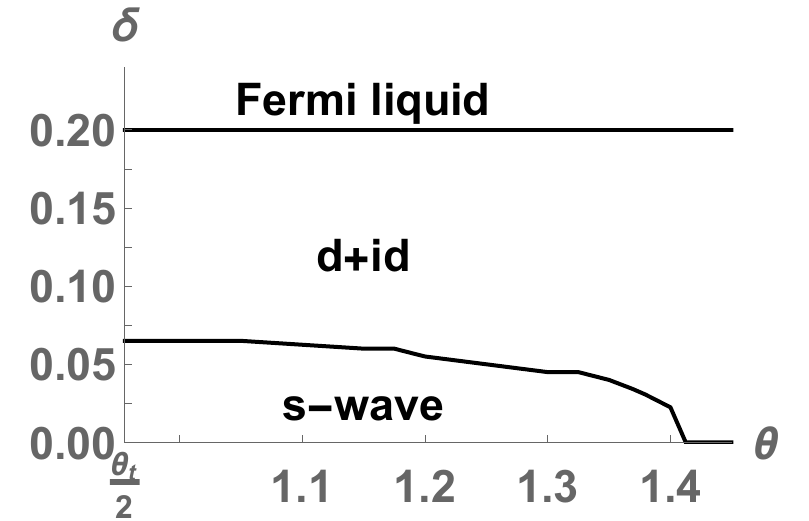}}
				\label{PD}
	}
	\subfloat[][KT spectra]{
				\includegraphics[height=3.6cm]{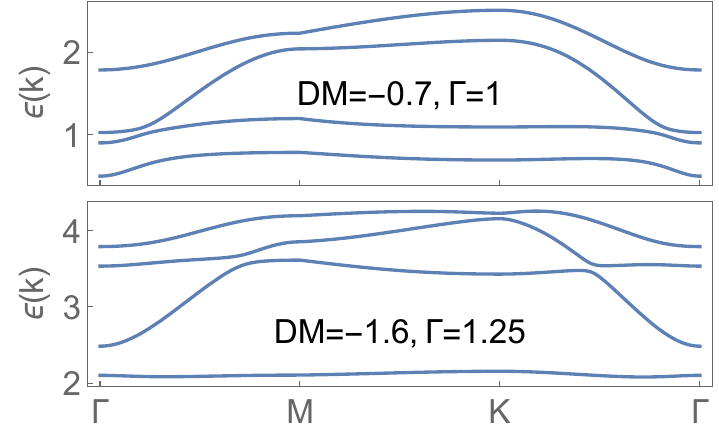}
				\label{MagnonKT}
				\newline
			}
			\caption{(color online) (a) Spin-wave spectra in the trilayer system for the DMI-dominated case $\theta =1.57$ and approaching the Heisenberg limit at $\theta \approx 0.96$. Energies are given in units of $t^2/U$ (b) Phase diagram for the t-J-like model in the trilayer system. (c) Spin-wave spectra in the bilayer system for a topologically non-trivial case $DM=-0.7$, and $DM=-1.6$, which is trivial. Energies are given in units of the Heisenberg coupling $J$.}
\end{figure*}

%%%%%%%%%%%%%%%%%%%%%%%%%%%%%%%%%%%%%%%%%%%%%%%
%\section{Topological magnon bands}
%%%%%%%%%%%%%%%%%%%%%%%%%%%%%%%%%%%%%%%%%%%%%%%
\emph{Topological magnon bands - }%
To study spinwaves in noncollinear magnetic structures, we first perform a sublattice dependent rotation to make the local $z$-axis point in the direction of the local moment, and then introduce a Holstein-Primakoff representation \cite{Auerbach}. The resulting Hamiltonian is truncated to quadratic order, Fourier transformed and then diagonalized using a Bogoliubov transform \cite{Maestro2004}. The magnon spectra for bulk pyrochlore iridates was studied in Ref.~[\onlinecite{PhysRevB.87.214416}], employing a parametrization in which	
\begin{align}
	J			&=	\frac{4t^2}{U} \left[ \cos^2 \left( \frac{\theta_t}{2}-\theta \right) - \frac{1}{3} \sin^2 \left( \frac{\theta_t}{2} - \theta \right) \right],\\
	\mathbf{D}_{ij}		&=	\frac{8t^2}{U} \cos \left( \frac{\theta_t}{2}-\theta \right) \sin \left( \frac{\theta_t}{2}-\theta \right) \hat{v}_{ij},\\
	\Gamma^{ab}_{ij}	&=	\frac{8t^2}{U} \sin^2 \left( \frac{\theta_t}{2}-\theta \right) \left[ \hat{v}^a_{ij} \hat{v}^b_{ij} - \frac{\delta^{ab}}{3} \right],
\end{align}
where $\theta_t = 2 \arctan \sqrt{2} \approx 109.47^\degree$ is the tetrahedral angle. In this parametrization it is clear that energies scale by $t^2/U$, and $\theta$ is the parameter determining the relative importance of the interactions. The system approaches the Heisenberg limit and becomes gapless as $\theta \rightarrow \theta_t /2 \approx 0.96$. In the bulk system, the cubic symmetry guarantees a threefold degeneracy at the $\Gamma$ point \cite{PhysRevB.87.214416}. This degeneracy is lifted in the thin film case, and we generically find an isolated lowest band, as can be seen in Figs.~\ref{MagnonTKT} and \ref{MagnonKT}, for the trilayer and bilayer systems, respectively. Since the lowest band is isolated, it has a well-defined (first) Chern number, which we calculate numerically using the Fukui method \cite{Fukui2005}. It turns out to be nonzero in a large portion of the parameter space, due to the noncoplanar spin texture, caused by the DM and $\Gamma$ terms, inducing a nontrivial Berry phase on magnon motion through the Brillouin zone. The relatively flat band also implies that magnon-magnon interactions could lead to correlated boson (magnon) behavior.

In the TKT spectra (Fig.~\ref{MagnonTKT}) we use the parametrization given above, and find that the direct gap between the two lowest bands 
is $0.114t^2/U$ ($0.135t^2/U$) for $\theta = 1.57$ ($\theta = 0.96$). 
Taking Y$_2$Ir$_2$O$_7$ as a typical example of a Mott-insulating AIAO magnet, we have the realistic value $U=2.5$eV from first-principles calculations on the bulk material \cite{Shinaoka2015}. Assuming $U/t=6$, the smaller gap is 
$92$K. 
Since the iridates are most likely in the intermediate coupling regime, the $U/t$ ratio is unlikely to be significantly higher. The gap should thus correspond to experimentally accessible temperatures 
away from a band closing at $\theta_c =1.26$. 
Defining the flatness ratio as band gap over bandwidth, the TKT bands (apart from the $\theta = 0.96$ case which approaches the Heisenberg limit) have flatness ratios on the order of $1/10$ \cite{Supplemental}. Throughout the parameter space considered, the lowest band has a nonzero Chern number $C=\pm 1$, with a topological transition at $\theta_c$. The full band evolution is shown in the Supplemental Material \cite{Supplemental}.

For the KT spectra (Fig.~\ref{MagnonKT}) we used values of $DM/J$ and $\Gamma/J$ chosen to engineer a state as close to the AIAO order as possible. Again, we stress that iridates are unlikely to reach the high DM strength required, but it might be possible to engineer in optical lattice systems. Supposing that is the case, we find that the gap closes around $DM/J=-1.3$, representing a topological transition from $C=-1$ at $|DM|<1.3$ to a topologically trivial system for $|DM|>1.3$ \cite{Supplemental}. In the topological sector, the gap is $0.3 J \approx 95$K, using $J=27.3(6)$meV as reported for Sm$_2$Ir$_2$O$_7$ \cite{PhysRevLett.117.037201}, giving a flatness ratio that is approximately one. In general we seem unable to achieve a both very flat and topologically nontrivial band in these systems, which would allow for interesting interaction effects \cite{PhysRevLett.116.066402}. Including longer-range interactions (which translate into longer range hopping terms for the spin waves) could perhaps change that, as in the case of electronic Chern insulators \cite{BERGHOLTZ2013}.

Since magnons are bosonic excitations, a quantized magnetic (thermal) Hall response is not expected. Instead we envision the kind of observations found in other topological bosonic systems \cite{PhysRevA.88.063631}, such as photonic topological insulators \cite{Khanikaev:nm13}, Hofstadter bands in ultracold bosonic atoms \cite{2015NatPh..11..162A}, topological bands in cold bosonic atoms \cite{PhysRevA.79.053639}, and topological polariton insulators \cite{PhysRevX.5.031001,PhysRevLett.114.116401}. Wave packets are excited into the nontrivial bands, possibly using a pulsed magnetic field as a pump, producing long-lived edge excitations. 

\begin{figure}[h]
	\subfloat[][TKT]{
	\includegraphics[height=2.75cm]{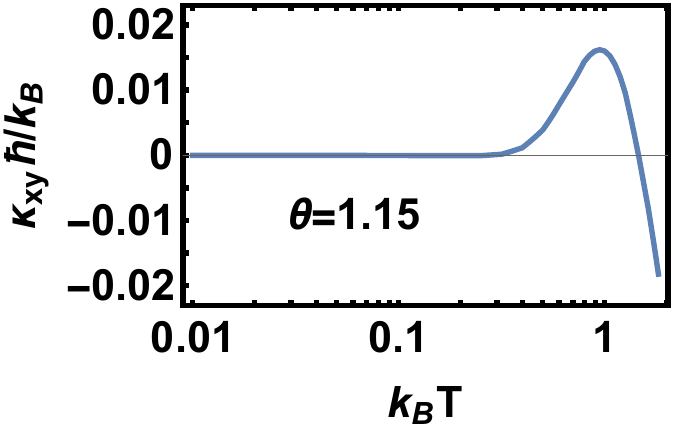}
	}
	\subfloat[][KT]{
	\includegraphics[height=2.75cm]{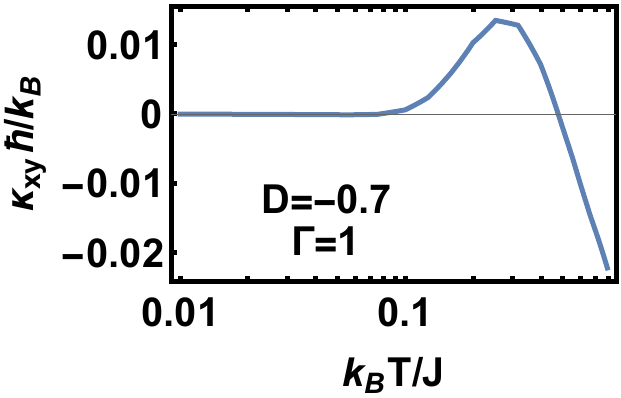}
	}
	\caption{\label{fig:Hall}Temperature dependence of the magnon Hall conductivity, $\kappa_{xy}(T)$ for (a) the TKT trilayer system and (b) the KT bilayer system, up to the estimated ordering temperature \cite{Supplemental}. The sign change of the thermal Hall conductivity with temperature is associated with a change in which topological bands dominate the transport \cite{Supplemental}.  In (a) $k_B T$ is given in units of $t^2/U$. }
\end{figure}

We have also computed (see Fig.\ref{fig:Hall}) a magnon Hall effect, in which the magnon edge current produces a thermal Hall current in the presence of a temperature gradient and an associated transverse thermal conductivity $\kappa_{xy}$ \cite{PhysRevLett.106.197202,PhysRevB.84.184406,PhysRevB.89.054420,PhysRevB.90.024412}. As a function of temperature, we find \cite{Supplemental} a sign change in $\kappa_{xy}$ that reflects the topological nature of the thermal Hall effect in this system as bands of different Chern number come to dominate the transverse thermal transport \cite{PhysRevB.91.125413}. 
Observing this sign change requires a sensitivity of $\lesssim 10^{-11}$ W/K \cite{Supplemental}.
 The thermal Hall effect has been experimentally demonstrated in collinear pyrochlore and kagome ferromagnets \cite{Onose2010,PhysRevB.85.134411,Hirschberger2015a,Hirschberger2015,PhysRevLett.115.147201,PhysRevB.89.134409}, and predicted to occur also in kagome antiferromagnets \cite{PhysRevB.90.035114,Owerre:cm16a,Owerre:cm16c}, as well as other systems \cite{0953-8984-28-38-386001,:/content/aip/journal/jap/120/4/10.1063/1.4959815,PhysRevLett.117.227201,Owerre:cm16b,Owerre:cm16}. The topology is due to the DMI in collinear systems, and finite spin chirality of the magnetic order in noncoplanar systems, which may be caused by DMI or magnetic fields. To our best knowledge, the current study is the first  to predict it will occur in this specific noncollinear and noncoplanar magnetic configuration. 
 The noncollinearity makes the order more susceptible to control by external magnetic fields.  For small fields we expect the thermal Hall conductivity should still be present since the lowest magnon band is separated by a gap from the next higher band.

%%%%%%%%%%%%%%%%%%%%%%%%%%%%%%%%%%%%%%%%%%%%%%%
%\section{Doping and superconductivity}
%%%%%%%%%%%%%%%%%%%%%%%%%%%%%%%%%%%%%%%%%%%%%%%
\emph{Doping and superconductivity - }%
To identify order parameters with the same symmetries as the spin Hamiltonian, we first employ a hidden symmetry \cite{PhysRevLett.69.836} to write the Hamiltonian, Eq.\eqref{Hmag}, in a Heisenberg form, $H=\sum_{ij} J_0 \mathbf{S}_i ' \cdot \mathbf{S}_j '$, where the site- and bond-dependent SO(3) transformation depends on the normalized DM vector $\hat{v}_{ij}$ according to,
\begin{align}
	\mathbf{S}_k'	&=	\left( 1-\cos \phi \right) \left( \hat{\mathbf{v}}_{ij} \cdot \mathbf{S}_k \right) \hat{\mathbf{v}}_{ij} + \cos \phi \mathbf{S}_k - \sin \phi_k \mathbf{S}_k \times \hat{\mathbf{v}}_{ij},
\end{align}
where $\phi_k = \phi \left( \delta_{ki}-\delta_{kj}\right)$. The angle $\phi$ can be related to the $\theta$ parameter used in the spin wave parametrization through $\phi = \theta_t/2 - \theta$. Using the methods of Refs.~[\onlinecite{PhysRevB.54.12946,PhysRevB.81.064428}], we next represent the rotated spin operators by fermionic spinons $f_\sigma$ through $S_i^{\prime a}	= \frac{1}{2} f_{i\sigma}^{\prime \dagger} \tau_{\sigma \sigma'}^a f_{i\sigma'}'$, subject to the local constraint \cite{RevModPhys.78.17} $\sum_\sigma f^{\prime \dagger}_{i\sigma} f'_{i\sigma}=1$, which will later be approximated by a global constraint for each sublattice. We identify the usual $t-J$ model order parameters for spin-conserving exchange, $\hat{\chi}'_{ij} =	f_{i\alpha}'^\dagger f'_{j\alpha}$ and singlet pairing, $\hat{\Delta}'_{ij} =	f'_{i\alpha} i\tau^y_{\alpha\beta} f'_{j\beta}	= f'_{i\uparrow} f'_{j\downarrow} - 	 f'_{i\downarrow}f'_{j\uparrow}$. These order parameters are then rotated back to the original coordinate system using a SU(2) transformation. This procedure gives us the parameters \cite{Supplemental}, 
\begin{align}
	\hat{\chi}_{ij}		=	f_{i\alpha}^\dagger f_{j\alpha},	\quad & \hat{\psi}_{ij}		=	f_{i\alpha}^\dagger \left( i \hat{v}_{ij} \cdot \vec{\tau} \right)_{\alpha \beta} f_{j\beta},	\\
	\hat{\Delta}_{ij}	=	f_{i\alpha} i\tau^y_{\alpha\beta} f_{j\beta},	\quad & \hat{\xi}_{ij}		=	f_{i\alpha} \left( \tau^y \hat{v}_{ij} \cdot \vec{\tau} \right)_{\alpha\beta} f_{j\beta},
 \end{align}
representing spin-conserving exchange, non-spin-conserving exchange, singlet pairing, and triplet pairing, respectively. Similar order parameters have been employed for Kitaev-Heisenberg models \cite{PhysRevB.87.064508,PhysRevLett.110.066403}, but in our model the components of the non-spin-conserving exchange and triplet pairing parameters get weighted by the DM vectors. This is a reflection of the spin-lattice coupling due to the the DM interaction, and the effect vanishes in the absence of spin-orbit coupling, as seen in the relations \cite{Supplemental},
\begin{align}
	\hat{\chi}'_{ij}	&=	\cos \left( \theta_t/2 - \theta \right) \hat{\chi}_{ij} + \sin \left( \theta_t/2 - \theta \right) \hat{\psi}_{ij},	\\
	\hat{\Delta}'_{ij}	&=	\cos \left( \theta_t/2 - \theta \right) \hat{\Delta}_{ij} - \sin \left( \theta_t/2 - \theta \right) \hat{\xi}_{ij},
 \end{align}
 where $\theta =\theta_t /2 $ for the Heisenberg case, as before.

Dopants are introduced through a hopping term $H_t = -t \sum_{\langle ij\rangle \sigma} \left( c_{i\sigma}^\dagger c_{j\sigma} + \mathrm{H.c.} \right)$, where $c^\dagger$ is an electron or hole creation operator (our treatment is symmetric in this respect). To disallow double occupation this fermion operator is represented as $c_{i\sigma}^\dagger =f_{i\sigma}^\dagger b_i$, where $b$ is the slave-boson operator subject to the constraint $1=b_i^\dagger b_i + \sum_\sigma f_{i\sigma}^\dagger f_{i\sigma}$. We assume that all bosons are condensed in their lowest band, i.e. $\delta =\langle b_i^\dagger b_i \rangle = \langle b_i^\dagger b_j \rangle$, which holds in the low-doping regime at zero temperature. This leads us to the mean-field hopping term $H_t = -t\delta \sum_{\langle ij \rangle} \left( f_{i\sigma}^\dagger f_{j\sigma} + \mathrm{H.c} \right)$. We further carry out mean-field decouplings for all our order parameters, assuming one value per sublattice. This results in a system of 36 mean-field parameters to be solved self-consistently \cite{Supplemental}. To make the problem more tractable, we introduce Ansatze for $s$- and time-reversal symmetry breaking $d$-wave states. In both cases we distinguish between bonds between sites both located in the kagome plane (in-plane bonds), and bonds between a site in the kagome plane and a site in a triangular layer (out-of-plane bonds). These are necessarily different, as in-plane and out-of-plane sites have different coordination numbers. We also take the exchange order parameters $\chi_{in}$, $\chi_{out}$, $\psi_{in}$ and $\psi_{out}$ to be real-valued and the triplet pairing order parameter to be zero. In the $s$-wave Ansatz, the singlet pairing parameter satisfies $\Delta = \left( \Delta_{in}, \Delta_{in}, \Delta_{out}, \Delta_{out} \right)$. In the $d+id$-wave Ansatz, $\Delta = \left( \Delta_{in}, \Delta_{in} e^{i2\pi/3}, \Delta_{out} e^{i\pi/3}, \Delta_{out} e^{i\pi/3} \right)$ \cite{Supplemental}.

The results of the self-consistent calculations are shown in Fig.~\ref{PD}. At low doping $\delta$ and spin-orbit coupling we find the $s$-wave Ansatz to be energetically favorable. There is a transition to the $d+id$-wave state at finite doping, the value of which decreases with increased spin-orbit strength. Over $\delta \approx 0.2$ the superconducting order parameter vanishes, leading to a Fermi liquid state. In both the $s$- and $d+id$-wave solutions the self-consistent results have the order parameter changing sign between in-plane and out-of-plane bonds, i.e. $\mathrm{sgn} \left( \Delta_{out} \right) = - \mathrm{sgn} \left( \Delta_{in}\right)$. This kind of sign change can be understood by analogy to layered superconductors, where noncentrosymmetricity can induce Rashba SOC with signs that alternate between layers, driving a similar sign change of the order parameter \cite{Sigrist2014,PhysRevB.84.184533,PhysRevB.92.174502}. While our model does not explicitly contain a Rashba-like term, spin-orbit physics is present in the form of DM and $\Gamma$ interactions. These, in turn, drive a magnetic order, the moments of which can be considered analogues of the effective magnetic field due to the Rashba effect. Considering the AIAO order of Fig.~\ref{MFT_TKT}, one finds that the in-plane moments can be averaged to an effective field $\mathbf{B}_{eff} \sim +\hat{z}$. Similarly the out-of-plane moments have $\mathbf{B}_{eff} \sim - \hat{z}$, which yields the same kind of sign structure.

%%%%%%%%%%%%%%%%%%%%%%%%%%%%%%%%%%%%%%%%%%%%%%%
%\section{Conclusion}
%%%%%%%%%%%%%%%%%%%%%%%%%%%%%%%%%%%%%%%%%%%%%%%
\emph{Conclusion - }%
In this Letter we have shown, using a strong coupling (local moment) model, that thin film pyrochlore iridate magnets are expected to exhibit topological magnon bands, with strong signatures in the temperature dependence of the thermal conductivity, as well as exotic time-reversal symmetry broken $d+id$ superconductivity upon doping. We focused on the ultrathin bilayer and trilayer systems grown in the [111] direction. Our work complements earlier studies that used the weak coupling limit as a starting point, and draws connections to layered superconductors and magnon Hall physics.

%\emph{Acknowledgements - }%
\begin{acknowledgments}
We thank Victor Chua, Bert Halperin, Rex Lundgren, and Allan H. Macdonald for helpful discussions.  We gratefully acknowledge funding from grants ARO grant W911NF-14-1-0579 and NSF DMR-1507621. This work used the Extreme Science and Engineering Discovery Environment (XSEDE), which is supported by National Science Foundation grant number ACI-1053575, and the Texas Advanced Computing Center (TACC).
\end{acknowledgments}

\bibliographystyle{apsrev4-1}
%\bibliography{../magnons,../pyro111}
%merlin.mbs apsrev4-1.bst 2010-07-25 4.21a (PWD, AO, DPC) hacked
%Control: key (0)
%Control: author (72) initials jnrlst
%Control: editor formatted (1) identically to author
%Control: production of article title (-1) disabled
%Control: page (0) single
%Control: year (1) truncated
%Control: production of eprint (0) enabled
%

\end{document}

\title{Supplementary material for Topological magnon bands and unconventional superconductivity in pyrochlore iridate thin films}
\author{Pontus Laurell}
\email{laurell@physics.utexas.edu}
\author{Gregory A. Fiete}
\affiliation{Department of Physics, The University of Texas at Austin, Austin, TX 78712, USA}
\date{\today}
\maketitle

\section{Coordinate systems and thin film DM vectors}
\begin{figure}[]
	\subfloat[][Bulk lattice]{
				 \includegraphics[height=3cm]{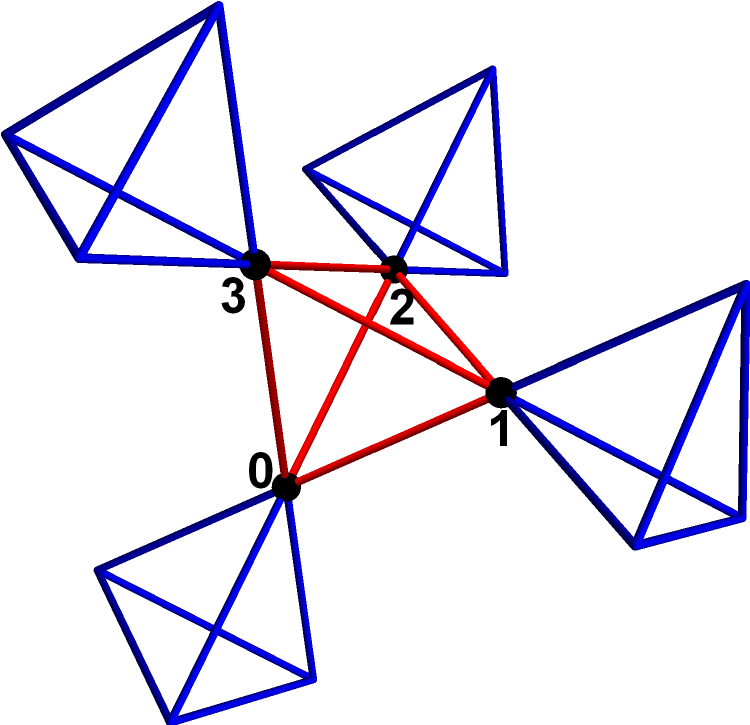}
				\label{BulkLattice}
	}
	\subfloat[][Bulk unit cell]{
				{\includegraphics[height=3cm]{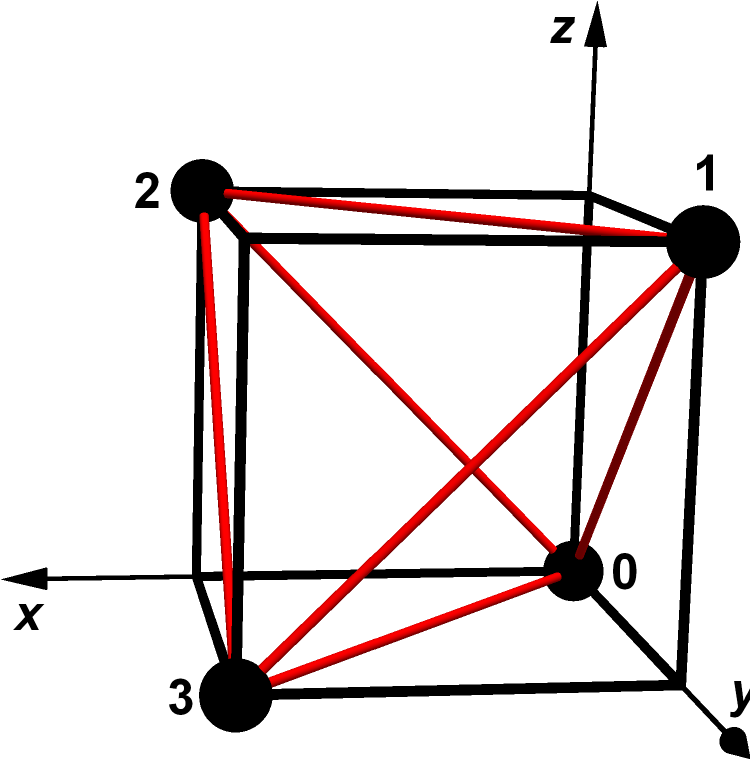}}
				\label{BulkUnitCell}
	}
			\caption{(color online) (a) Bulk pyrochlore lattice. Each tetrahedron vertex is the site of an Ir atom. The four sites in the tetrahedral unit cell are numbered $0$ to $3$. (b) The cubic coordinate system.}
\end{figure}

\begin{figure}
	\includegraphics[height=6cm]{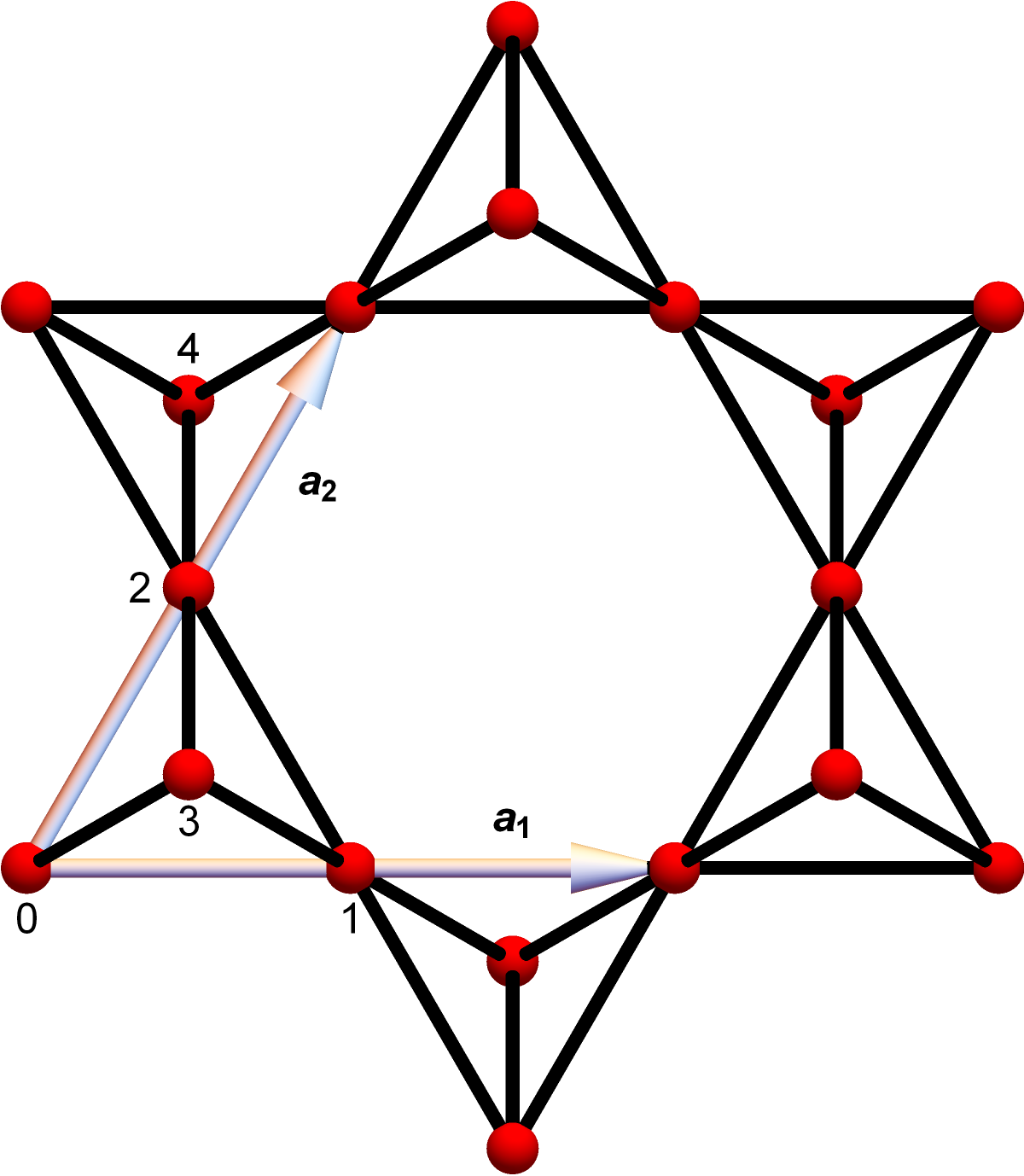}
	\caption{\label{TKTlattice}(color online) The thin film lattice structure. The five sites of the TKT trilayer unit cell are numbered $0$ to $4$. The KT bilayer has the same structure, but excludes the site labeled $4$. The lattice vectors $\mathbf{a}_1$ and $\mathbf{a}_2$ are also shown.}
\end{figure}

In this work, we neglect any physics due to the A site ions. When the A site is magnetic, the exchange between the rare-earth $f$ electrons and Ir $d$ electrons should in principle be taken into account. In practice, many magnetic phenomena appear independent of whether the A site is magnetic or not, suggesting that the Ir physics is most important \cite{PhysRevB.87.155101}. In addition, the $f-d$ exchange can often stabilize the Ir ordering \cite{PhysRevB.86.235129}. Our results will thus be directly applicable to pyrochlore iridates with nonmagnetic A ions, and are also expected to apply qualitatively to most cases with magnetic A ions.

We start with a bulk pyrochlore iridate in the cubic coordinate system described in figs.~\ref{BulkLattice} and \ref{BulkUnitCell}, in which each Ir atom in the tetrahedral unit cell resides on a vertex of a cube of length $a/(2\sqrt{2})$, where $a$ is the lattice constant \cite{PhysRevB.82.085111,PhysRevB.85.045124}. Next, we pick a kagome layer along the $[111]$ direction and rotate it into the $xy$-plane using
\begin{align}
	R_{bulk\rightarrow TKT}	&=	\left(\begin{array}{ccc}
												\frac{1}{\sqrt{2}} & \frac{1}{\sqrt{2}} & 0 \\
												 -\frac{1}{\sqrt{6}} & \frac{1}{\sqrt{6}} & \sqrt{\frac{2}{3}} \\
												 \frac{1}{\sqrt{3}} & -\frac{1}{\sqrt{3}} & \frac{1}{\sqrt{3}} \\
								\end{array}\right).
\end{align}
The resulting geometry is shown in fig.~\ref{TKTlattice}, having five (four) sites in the unit cell in the TKT trilayer (KT bilayer). The basis vectors are
\begin{align}
	\mathbf{r}_0				=	\left( 0,0,0 \right),	\quad  &\mathbf{r}_1				=	\frac{a}{2}	\left( 1,0,0 \right),	\quad 	\mathbf{r}_2				=	\frac{a}{4} \left( 1,\sqrt{3},0 \right),
\end{align}
\begin{align}
	\mathbf{r}_3				=	\frac{a}{4}	\left( 1, \frac{\sqrt{3}}{3}, \frac{2\sqrt{6}}{3} \right),	\quad &\mathbf{r}_4				=	\frac{a}{4}	\left( 1, \frac{5\sqrt{3}}{3}, -\frac{2\sqrt{6}}{3} \right),
\end{align}
with lattice vectors $\mathbf{a}_1	=	2\mathbf{r}_1$ and $\mathbf{a}_2	=	2\mathbf{r}_2$.

As alluded to in the main text, Moriya's rules \cite{PhysRev.120.91} allow determining the directions of the DM vectors up to a sign in bulk pyrochlore systems from symmetry \cite{PhysRevB.71.094420}. One uses the fact that a plane containing two Ir sites and the middle point of the opposite bond (the bond connecting the remaining two sites in the tetrahedron) is a mirror plane. By one of Moriya's rules, the DM vectors are perpendicular to this mirror plane, i.e. parallel to the opposite bond. In the thin film scenario, this is not possible in general. To see this, consider the plane containing any bond within the kagome plane in fig.~\ref{TKTlattice} and the center of the opposite bond. The associated mirror operation necessarily moves sites out of the plane. On the other hand, planes containing the bond between one site in the kagome plane and one site in a triangular layer do form mirror planes. Taken together, this means that the DM vectors for the thin film case are underdetermined if only symmetry is used. To overcome this difficulty, we assume that thin film systems inherit the bulk DM vectors. For the purpose of this paper, we consider a sandwich structure (see main text) with non-magnetic, lattice matched support layers. By viewing the DM vectors as a lattice property, the bulk DM vectors should be applicable throughout the heterostructure. In this picture, the difference between the magnetic Ir layer and the non-magnetic support layers is simply the absence of magnetic moments. The thin film DM vectors are thus defined to be
\begin{align}
	\tilde{v}_{\alpha\beta}^{rot}	&=	R_{bulk\rightarrow TKT} \tilde{v}_{\alpha\beta},
\end{align}
where $\alpha$ and $\beta$ are sublattice indices, and the six distinct normalized indirect bulk DM vectors are
\begin{align}
		\tilde{v}_{01}	= 	\frac{1}{\sqrt{2}} \left( 0,1,-1 \right),	\quad &	\tilde{v}_{02}=	\frac{1}{\sqrt{2}} \left( -1,0,1 \right),	\\
		\tilde{v}_{03}	=	\frac{1}{\sqrt{2}} \left( 1,-1,0 \right),	\quad &	\tilde{v}_{12}	=	\frac{1}{\sqrt{2}} \left( 1,1,0 \right),	\\
		\tilde{v}_{13}	=	\frac{1}{\sqrt{2}} \left( -1,0,-1 \right),	\quad &	\tilde{v}_{23}	=	\frac{1}{\sqrt{2}} \left( 0,1,1 \right).
\end{align}
Note that by convention, the pyrochlore DM vectors are given for the indirect configuration. Reversing the sign yields the direct configuration \cite{PhysRevB.71.094420}. The remaining DM vectors are related by antisymmetry, $\tilde{v}_{ji}=-\tilde{v}_{ij}$.

\section{Variational mean field calculations}
\begin{figure*}[]
		\subfloat[][Proximity to AIAO]{
			 \includegraphics[width=.4\linewidth]{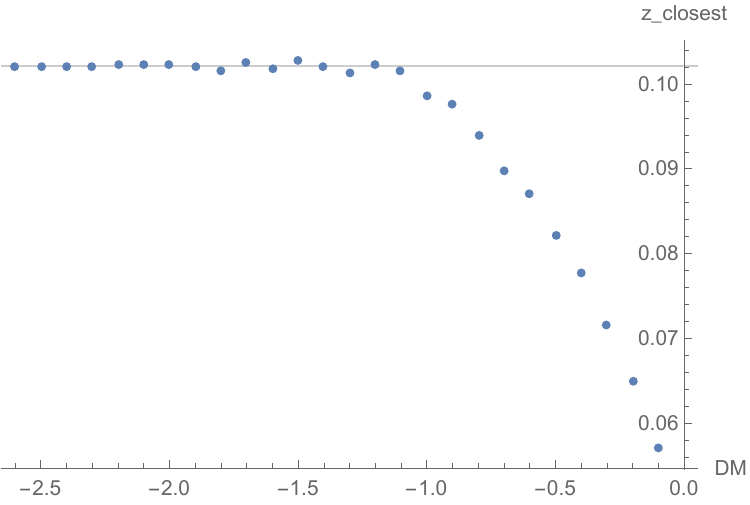}
			\label{KT_z_DM}
		}
		\subfloat[][$\Gamma (DM)$]{
			 \includegraphics[width=.4\linewidth]{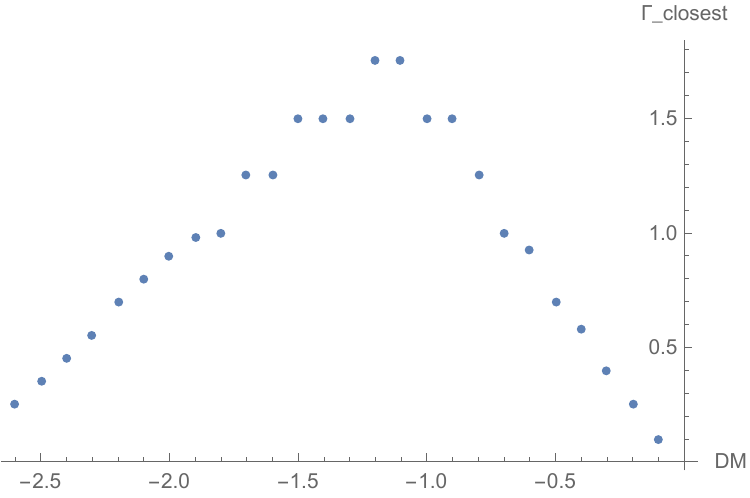}
			\label{KT_z_Gamma}
			\newline
		}
		\caption{(color online) (a) The plot shows how close the magnetic configuration of the KT bilayer system can get to the AIAO state (represented by the solid line) as a function of the DMI strength. (b) shows the value of the anisotropic coupling $\Gamma$ that realizes this approximate AIAO state, as a function of the DMI strength. The Heisenberg coupling is set to $J=1$.}
	\end{figure*}

Assuming a $\mathbf{q}=0$ structure with full ordering, $\langle \mathbf{S}_{\nu \beta} \rangle \equiv \langle \mathbf{S}_\beta \rangle$, the mean-field spin Hamiltonian can be written
\begin{align}
	H 	&=	\sum_{\mu\alpha} S_{\mu\alpha}^a g\mu_B B_{\mathrm{eff},\alpha}^a	\label{MFTthirdH}
\end{align}
where $\mu$ is a unit cell index, $\alpha$ a sublattice index and $a$ the vector index. The effective field (in units with $g\mu_B=1$) is
\begin{align}
		\frac{B_{\mathrm{eff},\alpha}^a}{2} 	&=	\sum_{\nu\beta} \left( J_{\mu\alpha\nu\beta} \langle S_\beta^a \rangle - \epsilon^{abc} D_{\mu\alpha\nu\beta}^b \langle S_\beta^c \rangle +  \Gamma_{\mu\alpha\nu\beta} \langle S_\beta^b \rangle\right).
\end{align}
Working with fixed length classical spins, we solve the model variationally, finding the spin configurations shown in the main text. The deformed AIAO states for the KT bilayer system are characterized by the $z$ coordinate where the spins intersect within the tetrahedron ($z=0$ in the kagome plane). In the AIAO order, the spins meet in the center of the tetrahedron, where $z=\left[ R_{bulk\rightarrow TKT} (1,1,1)/(4\sqrt{2}) \right]_z \approx 0.102$. As fig.~\ref{KT_z_DM} shows, we can get arbitrarily close to the AIAO configuration when the DM interaction dominates the Heisenberg coupling $J=1$. This also requires a particular choice for the anisotropic coupling $\Gamma$, given in fig.~\ref{KT_z_Gamma}.

The above treatment assumes that the sandwich structure is ideal, i.e. that the tetrahedra are not distorted. It is possible to consider the effect of weakly violating this assumption by scaling the exchange parameters on equivalent bonds. We have considered Ansatze in which (a) $J_{out}=c J_{in}$ with other exchange parameters unchanged, (b) scaling all out-of-plane parameters by $c$ (corresponding to scaling $t$ or $U$), and (c) $J_{out}=cJ_{in}$, $DM_{out}=cDM_{in}$ and $\Gamma_{out}=c^2 \Gamma_{in}$, where the superscript ${}_{in}$ (${}_{out}$) corresponds to in-plane (out-of-plane) bonds. We have explored the regime $c\in \left[ 0.7, 1.3 \right]$, $c\neq 1$, over which these Ansatze for both TKT and KT systems yield the same family of deformed AIAO states as previously found for the KT bilayers.

On these lattices, the Heisenberg exchange is associated with a large degeneracy. The degeneracy is lifted by the DMI, so the ordering temperature is expected to be set by the DM strength. In the related kagome \cite{PhysRevB.66.014422} (with DM vectors perpendicular to the plane) and bulk pyrochlore lattices \cite{PhysRevB.71.094420} (with direct DM vector configuration), the ordering temperatures follow $T_c \approx DM$, and $T_c \approx 2.5DM$, respectively. As a conservative estimate for ordering temperatures in the tri- and bilayer systems we scale the kagome $T_c$ by the ratio of the average coordination number to the coordination number of the kagome lattice, resulting in $T_c \approx 1.2DM$ ($T_c \approx 1.125DM$) for the TKT (KT) system.

\section{Spin wave calculation details}
The strong coupling Hamiltonian can be written
\begin{align}
    H	&=	\sum_{ij} S^a_i  \Lambda_{ij}^{ab}  S_j^b.
\end{align}
To describe deviations about the ground state, we rotate the spin quantization axis for each site such that $S_z$ points along the direction of the local moment,
\begin{align}
    S_i^a	&= \left[ R_i \left( \tilde{S}_i \right) \right]^a = R_i^{ab} \tilde{S}_i^b,
\end{align}
where $S_i^a$ is the spin operator with a global $z$ axis and $\tilde{S}_i^a$ has the axis unique to site $i$. Next we introduce the Holstein-Primakoff representation $\tilde{S}_i^z	=	s-a_i^\dagger a_i$,
\begin{align}
    \tilde{S}_i^+	=	\sqrt{2s-a_i^\dagger a_i} a_i,			\quad     &\tilde{S}_i^-	=	a_i^\dagger \sqrt{2s-a_i^\dagger a_i}.
\end{align}
While we have $s=1/2$ (from the $j_{eff}=1/2$ moments), a large-s expansion is still justified at low temperatures when $\langle a_i^\dagger a_i \rangle \ll 2s$. Truncating the expansion at linear order (neglecting magnon-magnon interactions), splitting the site index $i$ into a unit cell index $\mu$ and a sublattice index $\alpha$, Fourier transforming and carrying out a nearest-neighbor approximation results in the Hamiltonian
\begin{align}
	&H	=	\frac{1}{2}\sum_{\alpha\beta,}\sum_\mathbf{k} \left[ A_{\alpha\beta} \left( \mathbf{k}\right)	a_\alpha^\dagger \left( \mathbf{k} \right) a_\beta \left( \mathbf{k} \right) + B_{\alpha\beta} \left( \mathbf{k} \right) a_\alpha^\dagger \left( \mathbf{k} \right) a_\beta ^\dagger \left( -\mathbf{k} \right)	\right.\nonumber\\
		&+	\left. B_{\alpha\beta}^\star \left( -\mathbf{k}\right)	a_\alpha \left( -\mathbf{k} \right) a_\beta \left( \mathbf{k}\right) + A_{\alpha\beta}^\star \left( -\mathbf{k}\right) a_\alpha \left( -\mathbf{k} \right) a_\beta ^\dagger \left( -\mathbf{k}\right) \right],	\label{bilinearhamiltonian}
\end{align}
where
\begin{align}
	A_{\alpha\beta} \left( \mathbf{k}\right)	&=	\frac{1}{4} \left[ \tilde{\Lambda}_{\alpha\beta}^{xx} + \tilde{\Lambda}_{\alpha\beta}^{yy} -i\tilde{\Lambda}_{\alpha\beta}^{xy} + i\tilde{\Lambda}_{\alpha\beta}^{yx}\right] \mathcal{C}_{\alpha\beta} \left( \mathbf{k} \right)\nonumber\\
												&-\delta_{\alpha\beta} \sum_\gamma  \tilde{\Lambda}^{zz}_{\alpha\gamma} \mathcal{C}_{\alpha\beta} \left( \mathbf{0}\right),	\\
	B_{\alpha\beta} \left( \mathbf{k}\right)	&=	\frac{1}{4} \left[ \tilde{\Lambda}_{\alpha\beta}^{xx} - \tilde{\Lambda}_{\alpha\beta}^{yy} +i\tilde{\Lambda}_{\alpha\beta}^{xy} + i\tilde{\Lambda}_{\alpha\beta}^{yx}\right] \mathcal{C}_{\alpha\beta} \left( \mathbf{k}\right),
\end{align}
where, writing $\mathbf{r}_{\alpha\beta} = \mathbf{r}_\alpha - \mathbf{r}_\beta$,
\begin{align}
	\mathcal{C}_{\alpha\beta}  \left( \mathbf{k}\right)	&=	\left\{ \begin{array}{cc} 2\cos \left( \mathbf{k}\cdot \mathbf{r}_{\alpha\beta} \right) &	\quad,\,\alpha,\beta \in \{ 0,1,2\}\\ e^{i\left( \mathbf{k}\cdot \mathbf{r}_{\alpha\beta} \right)}	& \quad,\, \alpha \mathrm{~or~} \beta \in \{3,4\}\end{array}\right.,
\end{align}
% where
% \begin{align}
% 	A_{\alpha\beta} \left( \mathbf{k}\right)	&=	\frac{1}{2} \left[ \tilde{\Lambda}_{\alpha\beta}^{xx} + \tilde{\Lambda}_{\alpha\beta}^{yy} -i\tilde{\Lambda}_{\alpha\beta}^{xy} + i\tilde{\Lambda}_{\alpha\beta}^{yx}\right] \cos \left( i\mathbf{k}\cdot \mathbf{r}^{\alpha\beta} \right)\nonumber\\
% 												&-\frac{1}{2}\delta_{\alpha\beta} \sum_\gamma \left( \tilde{\Lambda}^{zz}_{\alpha\gamma} + \tilde{\Lambda}^{zz}_{\alpha\gamma} \right),	\\
% 	B_{\alpha\beta} \left( \mathbf{k}\right)	&=	\frac{1}{2} \left[ \tilde{\Lambda}_{\alpha\beta}^{xx} - \tilde{\Lambda}_{\alpha\beta}^{yy} +i\tilde{\Lambda}_{\alpha\beta}^{xy} + i\tilde{\Lambda}_{\alpha\beta}^{yx}\right] \cos \left( \mathbf{k}\cdot \mathbf{r}^{\alpha\beta}\right)
% \end{align}
and $\tilde{\Lambda}_{\alpha\beta}^{ab} =  \left[ R_\alpha ^T \Lambda_{\alpha\beta} R_\beta \right] ^{ab}$. The Hamiltonian is written on a matrix form by means of a Bogoliubov transformation,
\begin{align}
	H_\mathbf{k}	=	\mathbf{X}^\dagger h \mathbf{X},	\quad h \left( \mathbf{k} \right)	=	\left[ \begin{array}{cc} A \left( \mathbf{k} \right)	& B\left( \mathbf{k} \right) \\ B^\star \left(-\mathbf{k}\right)	& A^\star \left( -\mathbf{k} \right)\end{array}\right]
\end{align}
where $	\mathbf{X}	=	\left( \begin{array}{llllll}a_0\left(\mathbf{k}\right),	&	\dots	& a_m\left(\mathbf{k}\right),	& a^\dagger_0\left(-\mathbf{k}\right),	& \dots	& a^\dagger_m\left(-\mathbf{k}\right) \end{array}\right)^T$ and $m=4$ (3) for the TKT (KT) geometry. Since it is a bosonic system, extra care has to be taken to make sure that the Bogoliubov transformation is canonical \cite{Maestro2004,Colpa1978,Hemmen1980}. Thus the physical energies are the positive eigenvalues of $g h\left( \mathbf{k}\right)$, where
\begin{align}
	g	&=	\left( \begin{array}{cc}I_{n\times n}	&	0\\	0	& -I_{n\times n}\end{array}\right),
\end{align}
and $I_{n\times n}$ is the $n\times n$ identity matrix. For the TKT (KT) system, $n=5$ ($n=4$).

\begin{figure}
		\includegraphics[width=.8\linewidth]{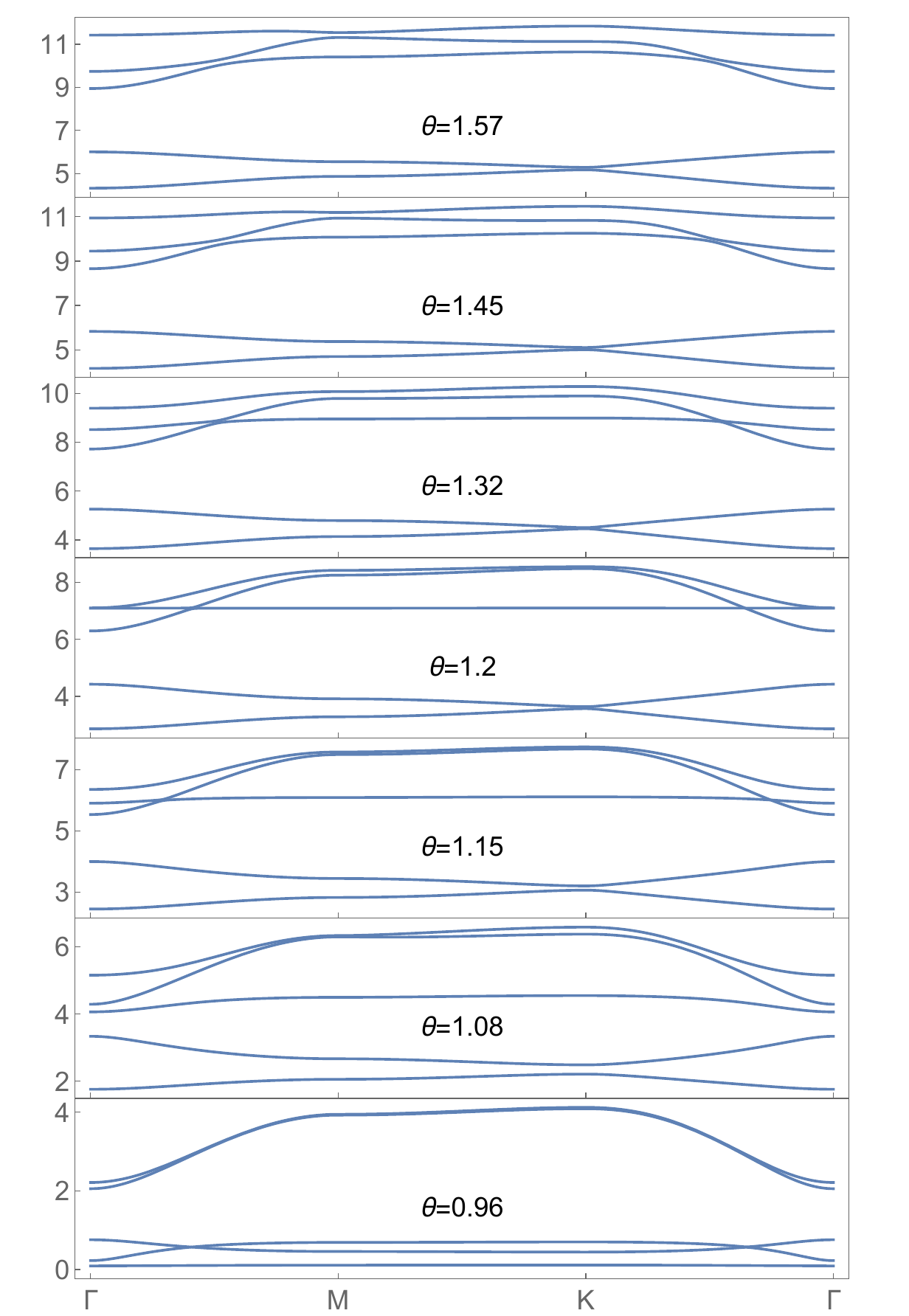}
		\caption{\label{MagnonTKTfull}(color online) Spin-wave spectra in the trilayer system. Energies are given in units of $t^2/U$. There is a band closing at $\theta_c =1.26$, associated with a sign change in the Chern number, from $+1$ for $\theta > \theta_c$ to $-1$ for $\theta < \theta_c$.}
	\end{figure}

\begin{figure}
		\includegraphics[width=.8\linewidth]{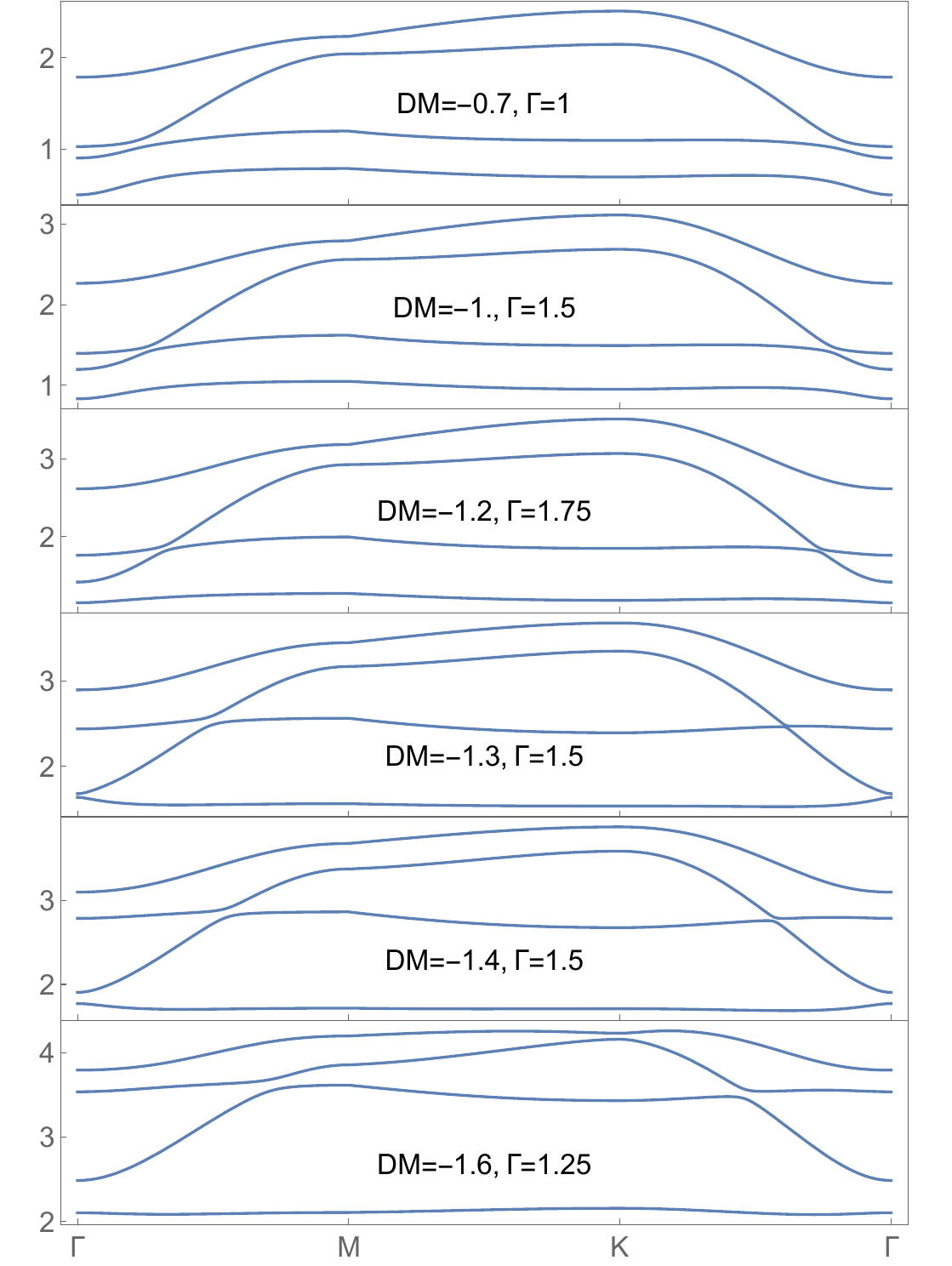}
		\caption{\label{MagnonKTfull}(color online) Spin-wave spectra in the bilayer system. Energies are given in units of the Heisenberg coupling $J$. There is a band closing around $DM=-1.3$, separating the topological sector with Chern number $-1$ from the trivial sector at $|DM|>1.3$.}
	\end{figure}

To complement the spectra shown in the main text, the full band evolutions are shown in figures~\ref{MagnonTKTfull} and \ref{MagnonKTfull}, for the TKT and KT systems, respectively.

\section{Thermal Hall conductivity}
One of the possible signatures of non-trivial Berry curvature in magnon bands is a thermal Hall effect for magnons. Here we calculate the thermal magnon Hall conductivity $\kappa_{xy}$ for both the TKT and KT geometries. We use the expression
\begin{align}
	\kappa_{xy}	&=	-\frac{k_B^2 T}{\hbar V} \sum_k \sum_{n=1}^N \left[ c_2 \left[ g\left( \epsilon_{nk} \right) \right] - \frac{\pi^2}{3}\right] \Omega_{n\mathbf{k}},	\label{kappaxy}
\end{align}
from Ref.~\cite{PhysRevB.89.054420}, which holds for general spin wave Hamiltonians that may have terms not conserving the total magnon number. Here $g\left( \epsilon_{nk} \right)$ is the Bose-Einstein distribution and $\Omega_{n\mathbf{k}}$ is the Berry curvature for the $n$:th band. The function $c_2$ is defined
\begin{align}
	c_2 \left( x \right)		&=	\left( 1+x \right) \left[ \ln \left( \frac{1+x}{x} \right) \right]^2 - \left[ \ln x \right]^2 - 2\mathrm{Li}_2 \left( -x \right),
\end{align}
where $\mathrm{Li}_n \left( x \right)$ is the polylogarithm, and has the limits 
$c_2 \left[ g\left( \epsilon_{nk} \right) \right] \rightarrow \pi^2/3$ ($ 0 $) as $\beta \epsilon \rightarrow 0$ ($\beta \epsilon \rightarrow \infty$). Hence, $\kappa_{xy}$ vanishes as $T\rightarrow 0$ and is generically non-zero for higher temperatures.

\begin{figure}
	\subfloat[][TKT, $\theta > \theta_c$]{
		\includegraphics[trim=0 0 0 0,height=2.8cm]{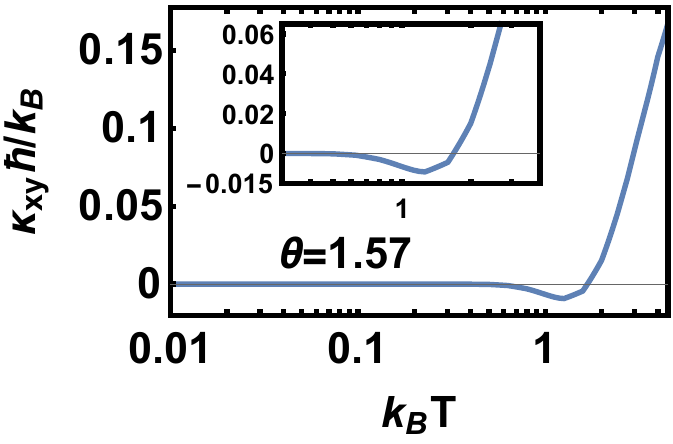}
		\label{KappaTKT_157}
	}%	
	\subfloat[][TKT, $\theta < \theta_c$]{
		\includegraphics[trim=0 0 0 0,height=2.8cm]{KappaTKT_115_Rescaled.pdf}
		\label{KappaTKT_096}
	}

	\subfloat[][KT, non-trivial sector]{
		\includegraphics[trim=0 0 0 0,height=2.8cm]{KappaKT_DM07_Rescaled.pdf}		
		\label{KappaKT_07}
	}%	
	\subfloat[][KT, trivial sector]{
		\includegraphics[trim=0 0 0 0,height=2.8cm]{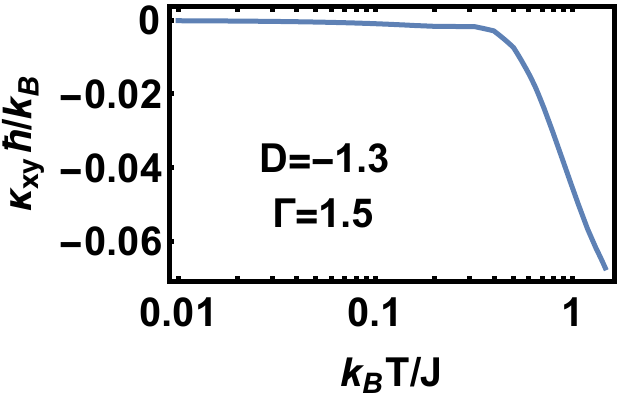}
		\label{KappaKT_13}
	}
	\caption{(color online) Magnon Hall conductivity for TKT trilayer system with (a) $\theta = 1.57$, and (b) $\theta=1.15$. The thermal energy $k_B T$ is given in units of $t^2/U$. In (c) and (d) magnon Hall conductivity is shown for the KT bilayer system with (c) $DM=-0.7$, and (d) $DM=-1.3$, respectively. In (a), (b) and (c), the sign of the initial bump depends on the sign of the Chern number of the lowest band. At higher temperatures, bands with opposite sign Berry curvatures contribute and eventually come to dominate. In (d) the system is in the trivial sector for the lowest band, and there is no sign change.\label{KappaPlots} }
\end{figure}

 We calculate $\Omega_{n\mathbf{k}}$ using the Fukui lattice discretization method \cite{Fukui2005} for momentum space grids of $1000\times 1000$ sites, and then compute $\kappa_{xy}$ for both the TKT and KT systems. The results are shown in fig.~\ref{KappaPlots}
 , for temperatures up to the estimated ordering temperatures. 
 The sign of the initial bump reflects the topological sector of the lowest band. We do observe a sign change in the thermal Hall conductivity as a function of temperature, due to the onset on the contribution of higher bands with opposite sign Berry curvatures. 
 For the KT systems in the topological sector and TKT systems with $\theta > 1.12$ the sign change occurs below the estimated ordering temperature. 
This sign change has also been shown to exist in ferromagnetic kagome systems \cite{PhysRevB.91.125413}, 
where it has also been shown to persist in the paramagnetic regime above the ordering temperature. We note, however, that it is not an ubiquitous feature. 
 In e.g. two-band models the Berry curvatures are necessarily equal and opposite, so the upper band can not overcome the magnitude of the Berry curvature of the lower band \cite{:/content/aip/journal/jap/120/4/10.1063/1.4959815}. 

In order to estimate the sensitivity required to experimentally resolve the sign change, we may use the magnitude of the initial bump. For TKT systems with $U=2.5$ eV, $t=U/6$, we find 
$2.2\cdot 10^{-11}$ W/K for $\theta =1.15$, and $1.7 \cdot 10^{-11}$ W/K for $\theta = 1.57$. For the KT bilayer with $J=27.3$ meV, we find $1.9\cdot 10^{-12}$ W/K.
%$6.7\cdot 10^{-10}$ W/K for $\theta =1.15$, and $8.8 \cdot 10^{-10}$ W/K for $\theta = 1.57$. For the KT bilayer with $J=27.3$ meV, we find $7.7\cdot 10^{-11}$ W/K. 
These values can be compared to experimentally achieved numbers for suspended thinfilms (less than $10^{-8}$ W/K, \cite{doi:10.1063/1.4704086}) and for nanowires ($10^{-11}$ W/K, \cite{doi:10.1063/1.3681255}). Since the conductance predicted here is a topological property, it should be possible to shrink and shape the sample in order to minimize spurious contributions from substrates etc.

The magnon bands may be broadened, particularly if the system is away from the strong-coupling limit. If two bands close to each other undergo such broadening, the higher band will contribute to the Hall conductivity at lower temperatures than in the non-broadened case. If the two bands have opposite Chern number, this leads to a suppression of the conductivity. When this happens to the two lowest bands, the initial peak in the Hall response will be reduced. However, while the Berry curvature is strongest near avoided crossings, there is also a non-negligible amount in other parts of the Brillouin zone. Hence the signal should only disappear if the bands come to overlap completely.

To estimate the broadening due to intermediate coupling we may look to the comparison of RPA and spin-wave calculations in bulk pyrochlore iridates \cite{PhysRevB.87.214416}. From their figures, the broadening can be estimated to be $\lessapprox 3$meV (or $35$K), at least away from the frustrated limit. Away from the topological transitions we thus expect only quantitative modifications to the magnon Hall response due to magnon broadening.

\section{Deriving order parameters}
In deriving the strong-coupling (spin) Hamiltonian used throughout this study, half-filling was assumed. In order to study doping effects and allow for deviations from half-filling, we reintroduce the density-density term by $H_{ij}' = H_{ij} - J n_i n_j/4$, where
\begin{align}
	H_{ij}	&=	J \mathbf{S}_i \cdot \mathbf{S}_j + \mathbf{D}_{ij} \cdot \left( \mathbf{S}_i \times \mathbf{S}_j \right) + S_i^a \Gamma_{ij}^{ab} S_j^b.
\end{align}
In order to define order parameters with the same symmetry as the Hamiltonian, we make use of a hidden symmetry due to Ref.~\cite{PhysRevLett.69.836} to write $H_{ij} = J_0 \mathbf{S}_i' \cdot \mathbf{S}_j'$, where
\begin{align}
		\mathbf{S}_i'	&=	\left( 1-\cos \phi \right) \left( \hat{\mathbf{v}}_{ij} \cdot \mathbf{S}_i \right) \hat{\mathbf{v}}_{ij} + \cos \phi \mathbf{S}_i - \sin \phi \mathbf{S}_i \times \hat{\mathbf{v}}_{ij},	\label{rotationSO3}\\
		\mathbf{S}_j'	&=	\left( 1-\cos \phi \right) \left( \hat{\mathbf{v}}_{ij} \cdot \mathbf{S}_j \right) \hat{\mathbf{v}}_{ij} + \cos \phi \mathbf{S}_j + \sin \phi \mathbf{S}_j \times \hat{\mathbf{v}}_{ij}.	\label{rotationSO3part2}
\end{align}
This hidden symmetry is a property of the spin Hamiltonian, and does not introduce any further assumptions. 
The above expressions describe rotations about the axis given by the DM vector $\hat{v}_{ij}$ by an angle $\phi_i = \phi = -\phi_j$. Using the identity
\begin{align}
		\left( \mathbf{A} \times \mathbf{B} \right) \cdot \left( \mathbf{C} \times \mathbf{D} \right)	&=	\left( \mathbf{A}\cdot \mathbf{C}\right) \left( \mathbf{B} \cdot \mathbf{D} \right) - \left( \mathbf{A}\cdot \mathbf{D}\right) \left( \mathbf{B} \cdot \mathbf{C} \right)
	\end{align}
for the scalar quadruple product, one finds
\begin{align}
		\mathbf{S}_i' \cdot \mathbf{S}_j'	&= \left[ \cos^2 \phi - \frac{\sin^2 \phi}{3} \right] \mathbf{S}_i \cdot \mathbf{S}_j + 2\sin \phi \cos \phi \hat{\mathbf{v}}_{ij} \cdot \left( \mathbf{S}_i \times \mathbf{S}_j \right)	\nonumber\\
											&+ 2\sin^2 \phi S_i^a \left[ v_{ij}^a v_{ij}^b - \frac{\delta^{ab}}{3} \right] S_j^b.
\end{align}
We identify the coefficients
\begin{align}
	J		&=	J_0	\left[ \cos^2 \phi - \frac{\sin^2 \phi}{3} \right],	\\
	DM		&=	J_0 2\sin \phi \cos \phi,\\
	\Gamma	&= J_0 2 \sin^2 \phi.
\end{align}
Comparing to the parametrization used in the main text we find $J_0 = 4t^2/U$ and $\phi = \theta_t/2 - \theta$.

In the rotated basis, the Hamiltonian
\begin{align}
	H_{ij}'	&=	J_0 \mathbf{S}_i' \cdot \mathbf{S}_j' - J n_i n_j/4
\end{align}
represents a standard $t-J$ model at $t=0$. Introducing a fermionic spinon representation $\mathbf{S}_i'=\frac{1}{2} f_{i\alpha}'{}^\dagger \vec{\tau}_{\alpha\beta} f_{i\beta}'$ (subject to the constraint $\sum_\sigma f_i^{\dagger \prime} f_i' = 1$), where $\vec{\tau}$ is the vector of Pauli matrices, we can immediately identify the standard order parameters $\hat{\chi}_{ij}'		=	f_{i\alpha}^{\dagger \prime} f_{j\alpha}'$ and $\hat{\Delta}_{ij}'	=	f_{i\alpha}' i\tau^y_{\alpha\beta} f_{j\beta}'$, for spin-conserving exchange and singlet pairing, respectively, and write
\begin{align}
	H_{ij}'	&=	\frac{J_0}{4} \left[ f_{i\alpha}^{\dagger \prime} f_{i\alpha}^\prime - \hat{\chi}_{ij}^\prime \left( \hat{\chi}_{ij}^\prime \right)^\dagger - \left( \hat{\Delta}_{ij}^\prime \right)^\dagger \hat{\Delta}_{ij}^\prime \right] - \frac{J}{4} n_i n_j.
\end{align}
To go back to the original basis, we note that the SO(3) rotations in eqs. \eqref{rotationSO3} and \eqref{rotationSO3part2} can also be implemented as a SU(2) rotation acting on the spinon degrees of freedom \cite{PhysRevB.54.12946}. Denoting the SO(3) rotation matrix by $R_{ij} \left( \phi \right)$ and the corresponding SU(2) transformation by $U_{ij}\left( \phi \right)$ we have
\begin{align}
		\mathbf{S}_i'	&=\frac{1}{2} f_{i\alpha}'{}^\dagger \tau_{\alpha\beta} f_{i\beta}' =  \frac{1}{2} f_{i\alpha}^\dagger R_{ij} \left( \phi \right) \vec{\tau}_{\alpha \beta} f_{i\beta}	\nonumber\\
				&=	\frac{1}{2} \left( f_i^\dagger U_{ij}^\dagger\left( \phi \right) \right)_\alpha \vec{\tau}_{\alpha\beta} \left( U_{ij}\left( \phi \right) f_i \right)_\beta,
\end{align}
i.e. $f_{i\alpha}'{}^\dagger	=	f_{i\beta}^\dagger \left[ U_{ij}^\dagger\left( \phi \right) \right]_{\beta\alpha}$. Applying this SU(2) rotation to the order parameters $\hat{\chi}'$ and $\hat{\Delta}'$ we find
\begin{align}
	\hat{\chi}'_{ij}	&=	\cos \left( \phi \right) \hat{\chi}_{ij} + \sin \left( \phi \right) \hat{\psi}_{ij},	\\
	\hat{\Delta}'_{ij}	&=	\cos \left( \phi \right) \hat{\Delta}_{ij} - \sin \left( \phi \right) \hat{\xi}_{ij},	
\end{align}
where
\begin{align}
	\hat{\chi}_{ij}		=	f_{i\alpha}^\dagger f_{j\alpha},	\quad 	&\hat{\psi}_{ij}		=	f_{i\alpha}^\dagger \left( i \hat{v}_{ij} \cdot \vec{\tau} \right)_{\alpha \beta} f_{j\beta},	\\
	\hat{\Delta}_{ij}	=	f_{i\alpha} i\tau^y_{\alpha\beta} f_{j\beta},	\quad	&	\hat{\xi}_{ij}		=	f_{i\alpha} \left( \tau^y \hat{v}_{ij} \cdot \vec{\tau} \right)_{\alpha\beta} f_{j\beta}.
\end{align}
In terms of the new order parameters, the $t-J$-like Hamiltonian becomes
\begin{align}
	H_{ij}^\prime	&=	-t \left( c_{i\sigma}^\dagger c_{j\sigma} + \mathrm{H.c.} \right) + \frac{J_0}{4} \left[ f_{i\sigma}^\dagger f_{i\sigma} - \cos^2 \phi \left( \hat{\chi}_{ij} \hat{\chi}_{ij} ^\dagger  + \hat{\Delta}_{ij}^\dagger \hat{\Delta}_{ij}  \right) \right.	\nonumber\\	
					&\left. - \sin \phi \cos \phi \left( \hat{\chi}_{ij} \hat{\psi}_{ij}^\dagger + \hat{\psi}_{ij} \hat{\chi}_{ij}^\dagger - \hat{\Delta}_{ij}^\dagger \hat{\xi}_{ij} - \hat{\xi}_{ij}^\dagger \hat{\Delta}_{ij} \right) \right.	\nonumber\\
					&\left. - \sin^2 \phi \left( \hat{\psi}_{ij} \hat{\psi}_{ij}^\dagger + \hat{\xi}_{ij}^\dagger \hat{\xi}_{ij} \right) \right]  - \frac{J}{4} n_i n_j ,
\end{align}
where we have also introduced the hopping term for the dopants, the $t$-term.

\section{Slave boson mean field theory}
To disallow double occupation, the fermionic dopant creation operator is represented as $c_{i\sigma}^\dagger = f_{i\sigma}^\dagger b_i$, where $b_i$ is a slave-boson subject to the constraint $1=b_i^\dagger b_i + \sum_\sigma f_{i\sigma}^\dagger f_{i\sigma}$. At low temperatures and doping, all bosons can be assumed to be condensed, so that $\langle b_i^\dagger b_i \rangle = \langle b_i^\dagger b_j \rangle = \delta$. This leads to the mean-field form for hopping term $H_t = - t\delta \sum_{\langle ij \rangle} \left( f_{i\sigma}^\dagger f_{j\sigma} + \mathrm{H.c.} \right)$ and the approximation $n_i n_j \approx 1 - b_i^\dagger b_i - b_j^\dagger b_j$.

We introduce mean-field decouplings for all the order parameters, assuming a translationally invariant solution. The result is a model with 36 parameters (9 a priori independent $\chi_{\alpha \beta}$ parameters etc.) to be solved self-consistently. To make it more tractable we introduce Ansatze for $s$- and $d$-wave superconductivity states that distinguish between in-plane bonds (bonds 01, 02 and 12 in fig.~\ref{TKTlattice}) and out-of-plane bonds (bonds 03, 04, 13, 14, 23, 24). In both ansatzes, the triplet pairing order parameters are assumed to be zero. The exchange parameters $\chi_{in}$, $\chi_{out}$, $\psi_{in}$, and $\psi_{out}$ are assumed to be real-valued and constant, with $\chi_{in}=\chi_{01}=\chi_{02}=\chi_{12}$ and so on. In the $s$-wave Ansatz $\left( \Delta_{01}, \Delta_{02}, \Delta_{03}, \Delta_{04} \right) = \left( \Delta_{in}, \Delta_{in}, \Delta_{out}, \Delta_{out}\right)$, and in the $d$-wave Ansatz $\left( \Delta_{01}, \Delta_{02}, \Delta_{03}, \Delta_{04} \right) = \left( \Delta_{in}, \Delta_{in}e^{i2\pi/3}, \Delta_{out}e^{i\pi/3}, \Delta_{out} e^{i\pi/3}\right)$. We also neglect $\sin^2 \phi$ terms, which is justified in the region of interest. The $d$-wave Ansatz is complex in momentum space, and thus breaks time reversal symmetry, and represents a $d+id$ state.

The constraint $1=\delta + \sum_\sigma f_{i\sigma}^\dagger f_{i\sigma}$ is relaxed to be satisfied on average in each sublattice, rather than on each site. To implement it, we use Lagrange multipliers $\lambda_{in}$ and $\lambda_{out}$ for in- and out-of-plane 
sublattices, respectively. At each iteration of the self-consistent solution we find values of the Lagrange multipliers such that the constraint is satisfied. Then the Hamiltonian is diagonalized, and we calculate expectation values for the order parameters using the expressions in the next section. The self-consistent parameters are then updated using conservative mixing, until convergence is found.
	
\section{Expectation values}
Throughout the self-consistent calculations, we want to evaluate expectation values of the order parameters derived above. The expressions for the expectation values are derived in a way similar to that of Ref.~\cite{PhysRevB.86.224417}. We rewrite the Hamiltonian
\begin{align}
		H-H_0	&=	\sum_k \mathbf{\varphi}_k^\dagger H_k \mathbf{\varphi}_k	=	\sum_k \mathbf{\varphi}_k^\dagger U_k D_k U_k^\dagger \mathbf{\varphi}_k	\\
			&\equiv \sum_k \mathbf{\gamma}_k^\dagger D_k \mathbf{\gamma}_k
\end{align}
where $H_0$ is the constant part, $D_k$ is a diagonal matrix with the eigenvalues of $H_k$ as its elements, $U_k$ is the matrix constructed from the (normalized) eigenvectors of $H_k$ and $\varphi_k$ is the vector
\begin{align}
		\varphi_{\mathbf{k}}			&=	\left( f_{0\uparrow}\left( \mathbf{k}\right),  f_{0\downarrow} \left( \mathbf{k}\right) \dots f_{4\downarrow}\left( \mathbf{k}\right),  f^\dagger_{0\uparrow} \left( -\mathbf{k}\right) , \right. \nonumber\\
	&\left. f^\dagger_{0\downarrow} \left( -\mathbf{k}\right)  \dots f^\dagger_{4\downarrow} \left( -\mathbf{k}\right) \right)^T,
\end{align}
introduced in the Bogoliubov transformation. Here $T$ denotes transpose. $\gamma_k$ represents the annihilation operator for the effective quasiparticle in the diagonal basis, and satisfies $\gamma_{k,a}		=	U^\dagger_{k,ab} \varphi_{k,b}$, where the vector indices $a,b$ run over the $20$ components of $\varphi$.

For clarity, summation over spin indices $\sigma$ is not assumed in the following. The expectation value for the density $n$ is calculated as by averaging over all sites on the same sublattice.
\begin{align}
	n_{\mu\alpha}	&=	\sum_\sigma \langle f^\dagger_{\mu\alpha\sigma}f_{\mu\alpha\sigma} \rangle	 =	\frac{N_{sublattices}}{N_{sites}} \sum_\mu \sum_\sigma \langle f^\dagger_{\mu\alpha\sigma}f_{\mu\alpha\sigma} \rangle	\nonumber\\
					&=	\frac{5}{N_{sites}} \sum_k \langle \varphi^\dagger_{k,2\alpha+1} \varphi_{k,2\alpha+1} + \varphi^\dagger_{k,2\alpha+2} \varphi_{k,2\alpha+2} \rangle	\nonumber\\
					&=	\frac{5}{N_{sites}} \sum_k \langle \gamma^\dagger_{k,b} \left( \left[ U_k^\dagger	 \right]_{b,2\alpha+1} \left[ U_k \right]_{2\alpha+1,c} \right. \nonumber\\
					&\left. + \left[ U_k^\dagger \right]_{b,2\alpha+2} \left[ U_k \right]_{2\alpha+2,c}\right) \gamma_{k,c} \rangle,
\end{align}
where $N_{sites}$ is the total number of lattice sites. The matrix element is non-zero if and only if $b=c$ and the state corresponding to $b$ is occupied, i.e. the $b$:th eigenvalue of $H_k$ is less than or equal to zero.  Using $\left[ U_k^\dagger \right]_{b,2\alpha+1}= \left[ U_k^\star \right]_{2\alpha+1,b}$ we write
 \begin{align}
		    n_{\mu\alpha}	&=	\frac{5}{N_{sites}} \sum_k \sum_{b~occ.} \left[ \left| \left[ U_k \right]_{2\alpha+1,b} \right| ^2 + \left| \left[ U_k \right]_{2\alpha+2,b} \right| ^2 \right]
\end{align}
Similarly, we calculate the expectation value for $\chi$
\begin{align}
	 \chi_{\mu \alpha ,\mu \alpha +v}	&=	\frac{5}{N_{sites}} \sum_k \sum_{b~occ.} e^{i k \cdot v} \left( \left[ U_k^\star \right]_{2\alpha +1,b} \left[ U_k \right]_{2\alpha_{\alpha +v} +1,b} \right.	\nonumber\\
							&+	\left. \left[ U_k^\star \right]_{2\alpha +2,b} \left[ U_k \right]_{2\alpha_{\alpha +v} +2,b} \right),
\end{align}
where $v$ denotes the displacement vector from $R_\mu + r_\alpha$ to some nearest neighbor and $\alpha_{\alpha+v}$ denotes the sublattice index of that site. The other order parameters have the expectation values
\begin{align}
		  \Delta_{\mu \alpha ,\mu \alpha +v}	&=	\frac{5}{N_{sites}} \sum_k e^{-i k \cdot v} \sum_{b~unocc.} \left( \left[ U_k \right]_{2\alpha +1,b} \left[ U_k^\star \right]_{10+2\alpha_{\alpha+v} + 2,b} \right.	\nonumber\\
							&- \left. \left[ U_k \right]_{2\alpha +2,b} \left[ U_k^\star \right]_{10+2\alpha_{\alpha+v} +1 ,b} \right),
\end{align}
\begin{align}
			\psi_{\mu\alpha ,\mu\alpha +v}	&=	\frac{5}{N_{sites}} \sum_k i e^{i k \cdot v} \sum_{b~occ.} \left\{ \left[ U_k^\star \right]_{2\alpha +1,b}\left[ U_k \right]_{2\alpha_{\alpha+v} +2,b} v^-_{\alpha ,\alpha_{\alpha+v}}  \right.	\nonumber\\
											&+ \left. \left[ U_k^\star \right]_{2\alpha+2,b} \left[ U_k \right]_{2\alpha_{\alpha+v}+1,b} v^+_{\alpha ,\alpha_{\alpha+v}}	\right.	\nonumber\\
											&+ \left. v^z_{\alpha ,\alpha_{\alpha+v}} \left( \left[ U_k^\star \right]_{2\alpha +1,b} \left[ U_k \right]_{2\alpha_{\alpha + v} +1,b} \right.\right.	\nonumber\\
											&-\left.\left. \left[ U_k^\star \right]_{2\alpha+2,b} \left[ U_k \right]_{2\alpha_{\alpha+v} +2,b} \right) \right\},
		\end{align}
where $v^\pm _{\alpha , \alpha_{\alpha +v}}	=	v^x_{\alpha , \alpha_{\alpha +v}} \pm i v^y_{\alpha , \alpha_{\alpha+v}}$, and
\begin{align}
			\xi_{\mu\alpha ,\mu \alpha +v}	&=	\frac{5}{N_{sites}} \sum_k i e^{-i k \cdot v} \sum_{b~unocc.} \nonumber\\
							& \left\{ \left[ U_k \right]_{2\alpha +1,b}\left[ U_k^\star \right]_{10+2\alpha_{\alpha+v} +1,b} \left( - v^+_{\alpha ,\alpha_{\alpha+v}} \right) \right.	\nonumber\\
											&+ \left. \left[ U_k \right]_{2\alpha+2,b} \left[ U_k^\star \right]_{10+2\alpha_{\alpha+v}+2,b} v^-_{\alpha ,\alpha_{\alpha+v}}	\right.	\nonumber\\
											&+ \left. v^z_{\alpha ,\alpha_{\alpha+v}} \left( \left[ U_k \right]_{2\alpha +1,b} \left[ U_k^\star \right]_{10+2\alpha_{\alpha + v} +2,b} \right.\right.	\nonumber\\
											&+\left.\left. \left[ U_k \right]_{2\alpha+2,b} \left[ U_k^\star \right]_{10+2\alpha_{\alpha+v} +1,b} \right) \right\}
\end{align}
Note that the pairing order parameters $\Delta$ and $\xi$ involve sums over unoccupied (positive energy) states, instead of occupied states.

\bibliographystyle{apsrev4-1}
%\bibliography{../magnonsSupp.bib}
%merlin.mbs apsrev4-1.bst 2010-07-25 4.21a (PWD, AO, DPC) hacked
%Control: key (0)
%Control: author (72) initials jnrlst
%Control: editor formatted (1) identically to author
%Control: production of article title (-1) disabled
%Control: page (0) single
%Control: year (1) truncated
%Control: production of eprint (0) enabled
%